\documentstyle[emulateapj,apjfonts,psfig,times]{article}
\def\CEW{EW(C\,{\sc iv})}
\def\OEW{EW([O\,{\sc iii}])}
\def\HEW{EW(He\,{\sc ii}~4686)}
\def\MO3{$M_{\rm [O~III]}$}
\def\IUE{{\it IUE}}
\def\HST{{\it HST}}
\def\ROSAT{{\it ROSAT}}
\def\ASCA{{\it ASCA}}
\def\SAX{{\it BeppoSAX}}
\def\kms{\ifmmode {\rm km~s}^{-1} \else km~s$^{-1}$\fi}
\def\ltsim{\raisebox{-.5ex}{$\;\stackrel{<}{\sim}\;$}}
\def\gtsim{\raisebox{-.5ex}{$\;\stackrel{>}{\sim}\;$}}
\def\Lya{\ifmmode {\rm Ly}\alpha \else Ly$\alpha$\fi}
\def\cii{C\,{\sc ii}}
\def\civ{\ifmmode {\rm C}\,{\sc iv} \else C\,{\sc iv}\fi}
\def\oiii{O\,{\sc iii}}
\def\nv{N\,{\sc v}}
\def\Hbeta{\ifmmode {\rm H}\beta \else H$\beta$\fi}
\def\siIV{Si\,{\sc iv}}
\def\siII{Si\,{\sc ii}}
\def\ovi{O\,{\sc vi}}
\def\neviii{Ne\,{\sc viii}}
\received{2001 September 28}
\begin{document}

\title{The Luminosity Dependence of UV Absorption in AGN}
 \author{Ari Laor}
\affil{Technion, Physics Dept., Haifa 32000, Israel.\\
laor@physics.technion.ac.il}

\author{W. N. Brandt}
\affil{The Pennsylvania State University, Dept. of
Astronomy \& Astrophysics, 525 Davey Lab, \\
University Park, PA 16802. niel@astro.psu.edu}

\begin{center}
To appear in 2002, ApJ 569 (April 20th).
\end{center}

\begin{abstract}
We describe the results of a survey of the UV absorption properties
of the Boroson \& Green sample of AGN, which extends from the Seyfert 
($M_V\simeq -21$) to the luminous quasar ($M_V\simeq -27$) level.
The survey is based mostly on  
\HST\ archival data available for $\gtsim 1/2$ of the 87 sample
objects. Our main result is that soft X-ray weak quasars (SXWQs,
10 AGN with $\alpha_{ox}\le -2$) show the strongest UV absorption
at a given luminosity, and their maximum outflow velocity, $v_{\rm max}$,
is strongly correlated with $M_V$ ($r_S=-0.95$).
This suggests that $v_{\rm max}$ is largely set by the
luminosity,
as expected for radiation-pressure driven outflows.
Luminous SXWQs have preferentially low [\oiii] luminosity, which 
suggests they are physically distinct from unabsorbed AGN, 
while non-SXWQs with UV absorption are consistent with being drawn
from the unabsorbed AGN population. We also find an indication that 
$v_{\rm max}/v_{\rm BLR}$ increases with
$L/L_{\rm Edd}$, as expected for radiation-pressure 
driven outflows. This relation and the  $v_{\rm max}$ vs. $M_V$
relation may indicate that the radiation-pressure force multiplier increases
with luminosity, and that
 the wind launching radius in non-SXWQs
is $\sim 10$ times larger than in SXWQs.

\end{abstract}


\section{INTRODUCTION}
Fast outflows are common in AGN.
The outflow properties are very different in high
and low-luminosity AGN. Whereas luminous high-$z$ 
quasars display outflows reaching a few $10^4$~km~s$^{-1}$ in 
$\sim 10$\% of the objects,
Seyfert galaxies display typical outflow velocities up to only
$\sim 10^3$~km~s$^{-1}$, but in $\sim 50$\% of the objects. 
{\em Why are the outflow properties so different at
low and high-luminosity?  How do the outflow properties
vary with luminosity? Are AGN with outflows normal AGN
seen at a preferred angle, or are they physically distinct 
objects?
How is the UV absorption related to X-ray absorption?}
The answers bear important clues to the origin
and acceleration mechanism of AGN outflows.

The largest and most systematic study of broad absorption line 
quasars (BALQs) was carried out by Weymann et al. (1991), using
ground-based spectroscopy of $z>1.5$ quasars. More recent 
well-defined samples of BALQs are emerging from other large 
ground-based surveys (Becker et al. 2000; Menou et al. 2001; 
Hall et al. 2001). However, these comprehensive studies are limited
to luminous quasars. Studies of absorption in low-luminosity
AGN are best done in low-$z$ objects and thus require \HST\
(see Crenshaw et al. 1999 for the most comprehensive study).
There are many detailed studies of various  
objects ranging in luminosity from Seyferts to quasars, but these
are focused on understanding the absorber properties of each object
and do not provide a clear comprehensive picture.

In this study we make a first step toward a
uniform and systematic survey of the UV absorption properties of AGN
from the Seyfert to the luminous quasar level. We primarily use archival
\HST\ observations
of the Boroson \& Green (1992; hereafter BG92) 
sample, which includes the 87 $z<0.5$ Palomar-Green 
(PG, Schmidt \& Green 1983) AGN, 
and extends from Seyferts at $M_V\simeq -21$ to luminous
quasars at $M_V\simeq -27$. The high-quality optical data set available
from BG92 allows us to explore relations between quasar
absorption and emission properties known as ``eigenvector 1''
(EV1; BG92; Boroson 2002),\footnote {EV1 is defined in BG92,
and represents a set
of correlated optical emission properties, including in particular the
strength of the [\oiii] and Fe~II emission, and the \Hbeta\ line
width.} which are also strongly correlated with X-ray and UV emission
properties (e.g., Laor et al. 1997, hereafter L97; Brandt \& Boller 1999;
Wills et al. 1999; Laor 2000). These correlations are particularly
interesting since they may be driven by fundamental parameters,
specifically the black hole mass and the accretion rate.

The UV absorption properties of some AGN from the BG92 sample have been
studied by various authors (see the references in Appendix A and in
Brandt, Laor, \& Wills 2000; hereafter BLW, \S\S 5, 6), and a complete census
of the \civ\ absorption equivalent width, \CEW, of all BG92 AGN with sufficient
quality UV spectra is provided by BLW. This study also provides a
complete systematic census of the effective 
optical-to-X-ray spectral slope,
$\alpha_{ox}$, based on \ROSAT\ data.\footnote{Note that more recent
\HST\ and \ASCA\ observations of some of the objects indicate that
in a few cases $\alpha_{ox}$ may vary significantly 
(e.g., Gallagher et al. 2001).} The $\alpha_{ox}$ was used by BLW to
define a complete sample of Soft X-ray Weak quasars (SXWQs), which
includes all 10 AGN from BG92 with $\alpha_{ox}\le -2$. The
$\alpha_{ox}\le -2$ cutoff is based on the distribution of
 $\alpha_{ox}$ in the BG92 sample, which suggests an apparently
 distinct group of SXWQs (see BLW). Since
 $\langle \alpha_{ox}\rangle\simeq -1.48$ in non-SXWQs
 (Laor et al. 1997, and \S 3.2 here), the 2~keV luminosity of SXWQs is
 suppressed by a factor of $\ge 25$ on average, compared to non-SXWQs
 with the same optical luminosity. BLW
discovered a very strong relation between $\alpha_{ox}$ and \CEW,
which suggests that soft X-ray weakness is due to absorption,
as was directly confirmed for some objects by hard X-ray spectroscopy
(e.g., Gallagher et al. 1999, 2001; Green et al. 2001).

The \CEW\ of the 10 SXWQs covers a wide range ($1-100$~\AA), and
although nearly all BALQs are SXWQs (e.g., Kopko, Turnshek, \& Espey 1994;
Green et al. 1995; Green \& Mathur 1996), the
converse is not true as only some of these SXWQs can be defined
as BALQs (see below).
{\em What determines if a SXWQ is a BALQ?}
The purpose of this paper is to answer this question, together with
some of the questions posed in the opening paragraph. The paper is
organized as follows. In \S 2 we
briefly describe the method of analysis, and in \S 3 we provide
the absorption statistics and discuss their dependence on \MO3,
$M_V$, radio-loudness, emission-line strengths, and fundamental
parameters. Some further open questions are discussed in
\S 4, and the main conclusions are summarized in \S 5. Appenix A
provides notes on individual objects, supplementing the notes in
BLW, and Appendix B lists the 28 objects without intrinsic UV
absorption.

\section{Method}

We retrieved all public Faint Object Spectrograph and
High Resolution Spectrograph spectra of the BG92 AGN 
available in the \HST\ archives, and supplemented these with
the four \IUE\ spectra of SXWQs presented by BLW. 
UV coverage of \civ\ and/or \Lya+\nv\ was obtained for 56 objects
which extend from $M_V=-21.44$ to $M_V=-27.26$.  
Of these, 42 have \HST\ coverage of both \civ\ and 
\Lya+\nv, and six have only \civ\ coverage. Five objects
have no \civ\ coverage, and
\Lya+\nv\ was used instead. In three additional objects (SXWQs) 
both \civ\ and \Lya+\nv\ were obtained from \IUE.

The wavelength scale of each spectrum has been calibrated using
low-ionization Galactic absorption lines (O~I, C~II, Si~II, Al~II)
which are detectable in most spectra (e.g., Savage et al. 2000). 
We measure the intrinsic UV absorption using the \civ\
line since it is generally a strong absorption line, 
because it is relatively
simple to define the emission shape in its vicinity (unlike \nv), and 
because it is much less likely than \Lya\ to be affected by 
intervening systems. The absorption is parametrized here using three
numbers, the absorption rest-frame equivalent width, \CEW, 
the maximum velocity of absorption, $v_{\rm max}$ (expected to be
physically related to luminosity, see \S 3.4), and the velocity of
 maximum 
absorption, $v_{\rm \tau,max}$ (much less sensitive to continuum 
placement than $v_{\rm max}$). 
In some objects the \civ\ absorption appears 
only marginally significant. To verify its existence we required
that the second \civ\ doublet component be present (if the absorption
is narrow enough), and that there is
a feature at the same velocity in either \nv\ or \Lya. 
In three of the five objects where \civ\ is not available, absorption is
detected in \Lya\ and the values for \Lya\ are reported here. 
For the sake of brevity we use the term ``\civ\ absorption'' throughout
the paper, but we use a special symbol to denote the three
\Lya\ objects in the figures. 

Absorption was searched for from 0~\kms\ to 
30,000~\kms\ in blueshift with respect to the systemic
redshift determined from
[\oiii]~$\lambda 5007$ (provided by T. A. Boroson 2001, 
private communication). At velocities higher than 30,000~\kms\
the \civ\ absorption may be confused with \siIV\ absorption,
and for the sake of simplicity we avoid this velocity range.
The systemic redshifts are not heliocentric,
but the resulting error is smaller than the typical spectral 
sampling element size of $\sim 50-100$~\kms. The mean heliocentric 
velocity of the low-ionization Galactic absorption lines, used here
to calibrate the wavelength scale of the
\HST\ spectra, is also typically well below 100~\kms\ (e.g., 
Savage et al. 2000), and thus we expect our \HST\ velocity calibration 
to be accurate to a level of $\sim 100$~\kms, or better (excluding the 
four objects were no Galactic absorption was identified, as noted in 
Table~1). 

The absorption EW is measured by making a linear interpolation
(in log $f_{\lambda}$ vs. log $\lambda$) for the unabsorbed flux between
the two closest continuum points judged to be unabsorbed.
There is no accurate way to estimate the EW uncertainty, as it is
most likely dominated by systematic errors in the continuum 
placement. As a rough guide we estimate the associated error is
typically $\sim 10-20$\%, and generally
not below $\sim 0.1$~\AA. Very shallow (depth$< 10$\%) and
broad (width$>$ a few thousand~km~s$^{-1}$) absorption troughs
would remain undetected in our sample.

\section{Results \& Discussion}

\subsection{Absorption Frequency}

Absorption is detected in 28 of the 56 BG92 sample objects, and their
\CEW, $v_{\rm max}$, and $v_{\rm \tau,max}$ values are listed in Table~1.
In 21 objects the parameters are based on \HST\ \civ\
profiles, in three on \HST\ \Lya\ profiles, and in four on \IUE\ \civ\ 
profiles (all SXWQs from BLW). In two of these last four the absorption 
is only marginally significant (PG~1011$-$040 and PG~2214+139, see 
Appendix A), and we use a special symbol to denote them in the figures. 
Table~1 also lists the redshift, the 3000~\AA\ to 2~keV spectral 
slope, $\alpha_{ox}$, taken from BLW, $M_V$ and \OEW\ from BG92, and
the continuum luminosity at 3000~\AA, based on the continuum fluxes
from Neugebauer et al. (1987). A note to the table provides the
Balnicity Index (BI), as defined by Weymann et al. (1991), for the
five objects where BI$>0$.

Strong absorption, defined here as \CEW$>$10~\AA, is detected in
6/56 of the sample objects, of which five have BI$>0$,
i.e. a frequency of $\sim 10$\% for BALQs/strong absorption AGN.
Intermediate-strength absorption, 1~\AA$<$\CEW$<$10~\AA, is detected 
in 11/56, $\sim 20$\%, and weak absorption, 0.1~\AA$<$\CEW$<$1~\AA,
in 11/56, $\sim 20$\%. The remaining 50\% have no detectable
absorption, i.e. \CEW$\ltsim $0.1~\AA, and they are listed in
Appendix B. However, these frequencies may
be subject to significant systematic errors since the available 
subsample of 56 BG92 AGN is not complete or well-defined in any
way, and some of the objects in it were observed because
of their absorption properties.

Figure~1 shows the absorption profiles for the 10 SXWQs.
Figure~2 shows the absorption profiles for the 11 non-SXWQs
with intermediate UV absorption, and Figure~3 shows the 11 objects
with weak absorption.

\subsection{\CEW\ vs. $\alpha_{ox}$}

Figure~4 presents a revised version of Fig.~4 from
BLW, as some of the \CEW\ values found here are somewhat
different from those listed in BLW (see Appendix A).
As discussed in BLW, all the SXWQs have intermediate-to-strong UV 
absorption (with possible exceptions for
PG~1011$-$040 and PG~2214+139), and all the objects 
with strong UV absorption are SXW, consistent with the earlier
result that practically all BALQs are SXW. 

The second notable group is the non-SXWQs with 
intermediate UV absorption (marked by stars in Fig.~4). 
The seven objects in this group are also intermediate in their
$\alpha_{ox}$, having $\langle\alpha_{ox}\rangle=-1.68\pm 0.11$.  
They have almost no overlap with the 
$\alpha_{ox}$ of the 11 objects with weak UV absorption, which
have $\langle\alpha_{ox}\rangle=-1.49\pm 0.08$.
As noted in BLW, X-ray spectra available for some of these
intermediate objects suggest intrinsic X-ray absorption features as well, 
but the S/N is generally too low to identify the features (excluding
PG~1114+445 which shows what appears to be
a warm absorber edge, George et al. 1997).
These intermediate objects are relatively X-ray bright and would be
interesting to explore with higher resolution X-ray 
spectroscopy.

Nearly all objects with
no UV absorption have a ``normal'' $\alpha_{ox}$ $\sim -1.5$.
The two significant exceptions are PG~1259+593 and PG~1543+489, which
have significantly steeper $\alpha_{ox}$. Interestingly, these two 
objects also have very unusual UV line profiles (very broad, highly 
blueshifted, and very asymmetric), which may be related to their
unusual $\alpha_{ox}$.

\subsection{The \MO3\ distribution}

As pointed out by BLW (see Fig.~3 there), SXWQs tend to lie toward the
weak-[\oiii] end of the BG92 EV1 (see \S 1).
This result is consistent with the findings of Boroson \& Meyers (1992)
that low-ionization BALQs are much more common in low-\OEW\ AGN,
and the similar finding of Turnshek et al. (1997) for normal BALQs,
as both of these types of BALQs are generally SXW. BLW further showed
that the difference in the [\oiii] luminosities of SXW AGN vs.
non-SXW AGN is more significant than the difference in their
\OEW. Are non-SXW
AGN with UV absorption also more common at low [\oiii] luminosities?
We first take a closer look at the [\oiii] luminosities of SXWQs and
then address this question (the difference in \OEW\ is discussed
in \S 3.5).

Figure~5 shows the 
positions of the 56 AGN studied here in the \MO3\ vs. $M_V$ plane,
where \MO3$\equiv M_V-2.5\log~$\OEW\ (defined in BG92) is used
here as a measure of the [\oiii] luminosity.
Radio-loud and radio-quiet
AGN are shown separately in Fig.~5 since their \MO3\ and 
$M_V$ distributions are significantly different (see Table~5 in BG92).
All 9 radio-quiet SXWQs lie at \MO3$>-27$, where they form
$\sim 35$\% (9/26) of the population. The Poisson probability that the 
true fraction of SXWQs at \MO3$<-27$ is also $\sim 35$\% is
$7.8\times 10^{-3}$. PG~1004+130 is the only SXWQ of the 14 radio-loud
AGN in our sample, and it also lies at the bottom of the radio-loud
\MO3\ distribution. A KS test for the combined sample of SXWQs gives a 
probability of 1.5\% that the 10 SXWQs and the 28 AGN with no UV 
absorption are drawn from the same \MO3\ distribution. In this combined
sample the fraction of SXWQs at \MO3$<-27$ is $\sim 3$\% (1/30)
(the single object is PG~1004+130),
and the Poisson probability that the true fraction
is $\sim 35$\%, as found for \MO3$>-27$, is $3.2\times 10^{-4}$. 

Inspection of Fig.~5 suggests that the difference in the
\MO3\ distributions of SXWQs and non-SXWQs is mostly driven by the 
more luminous AGN. Indeed, a KS test for the complete BG92 sample of
87 AGN gives a probability of 1.2\% that the 4 SXWQs and 20 
non-SXWQs more luminous than $M_V=-25$ are drawn from the same
\MO3\ distribution, while for AGN less luminous than
$M_V=-25$ the probability for that is 18\%
(6 SXWQs and 57 non-SXWQs). 

In contrast to SXWQs, non-SXWQs with intermediate and weak UV absorption
are not significantly different in their \MO3\ distribution
from unabsorbed AGN (see Fig.~5). The KS test gives a probability of 
39\% that intermediately-absorbed and unabsorbed AGN are drawn from
the same distribution, and a corresponding probability of
97\% for the weakly absorbed AGN. {\em What do these}
\MO3\ {\em distributions imply?}

The [\oiii] emission is most likely isotropic
(e.g., Kuraszkiewicz et al. 2000 and references therein). Thus, as BLW 
pointed out (\S 8 there), the weakness of [\oiii] in SXWQs
strongly suggests they are physically distinct from normal 
$\alpha_{ox}$ quasars, as suggested by Boroson \& Meyers (1992)
and Turnshek et al. (1997) for low-ionization and normal BALQs.
The distributions in Fig.~5 suggest then that SXWQs, in particular
the luminous ones, 
are physically distinct from AGN with no UV absorption, but that 
AGN with intermediate to weak UV 
absorption are consistent with being drawn from the unabsorbed AGN 
population.

The low [\oiii] luminosities of luminous SXWQs argue then against
the common interpretation in which BALQs are normal quasars seen edge 
on (e.g., Weymann et al. 1991). An alternative
interpretation is that there is a range of
covering factors of the obscuring material
in AGN, and this covering factor strongly influences \MO3. For example,
the numbers found in 
our incomplete (and possibly biased) study would suggest that in AGN
with \MO3$<-27$ $\sim 57$\% of the ``sky'' is clear,
$\sim 20$\% is covered by low-column density clouds,
$\sim 20$\%  by intermediate-column density clouds, and only $\sim 3$\%
by high-column density clouds producing strong UV and X-ray absorption,
while in \MO3$>-27$ AGN $\sim 40$\%  of the ``sky'' is
clear, $\sim 25$\% is covered by low-to-intermediate
column density clouds,
and $\sim 35$\% is covered by high-column density clouds (see
Table~2). 
Note that the clouds' covering factors may be underestimated 
if the optical continuum emission of UV absorbed quasars is 
suppressed (Goodrich 1997; Krolik \& Voit 1998), as suggested by
the recent composite spectra presented by Brotherton et al. (2001).

\subsection{The Luminosity Dependence}

What determines if a SXWQ is a BALQ?
The answer appears to be remarkably simple. Figure~6 shows the
distribution of \CEW\ and of $v_{\rm max}$ vs. $M_V$ for all
28 absorbed AGN. SXWQs have a higher \CEW\ and a higher
$v_{\rm max}$, at any given luminosity, than intermediate and weak 
UV absorption AGN. In addition, both \CEW\ and $v_{\rm max}$ of the SXWQs 
are strongly correlated with $M_V$. The 
Spearman rank order correlation coefficient, 
$r_S$, and its significance level, Pr, for the \CEW\
vs. $M_V$ correlation are $-0.93$ and $1.1\times 10^{-4}$, and
for the $v_{\rm max}$ vs. $M_V$ correlation they are
$-0.95$ and $2.3\times 10^{-5}$. A least-squares fit to the SXWQs
gives \CEW$\propto L^{0.68\pm 0.18}$, and 
$v_{\rm max}\propto L^{0.62\pm 0.08}$. Only five objects in our
sample are ``true'' BALQs, using the Weymann et al. (1991) definition
of BI$>0$ (see the note to Table~1), and these are also the five most
luminous SXWQs. Thus, BALQs are simply luminous SXWQs.

On the other hand, for non-SXWQs there is no significant luminosity 
dependence of either \CEW\ or $v_{\rm max}$ (there may be a negative 
trend of \CEW\ with $M_V$ for the low-\CEW\ AGN, but the 
significance is $<2\sigma$). The differences in the $M_V$ dependence 
for SXWQs and non-SXWQs suggests that the UV absorption in these two 
populations occurs in physically different classes of absorbers.

The strong luminosity dependence of $v_{\rm max}$ is consistent
with some radiation-pressure acceleration scenarios. For example,
if the BAL outflow is launched by radiation-pressure acceleration
on optically thin clouds moving at some Keplerian
velocity $v_{\rm Kep}$, at some radius $R$, then 
the terminal outflow velocity is
\[ v_{\rm ter}\simeq v_{\rm Kep}\times \sqrt{\Gamma L/L_{\rm Edd}}, \]
where $\Gamma$ is the force multiplier, which depends on the 
opacity sources in the gas (dust, lines, or bound-free absorption), and is
assumed to remain constant throughout the flow. Since
\[ v_{\rm Kep}\propto \sqrt{M_{\rm BH}/R}\ \ \ {\rm and}\ \ \
L_{\rm Edd} \propto M_{\rm BH} \]
we get that 
\[ v_{\rm ter}\propto \sqrt{\Gamma L/R} .\]
Thus, in an object with a given $L$, the maximum terminal outflow velocity
is obtained for a flow which starts at the minimum launching radius,
$v_{\rm max}=v_{\rm ter}(R_{\rm min})$. If the main opacity
source is dust (Scoville \& Norman 1995), then the minimum radius 
where grains survive is
\[ R_{\rm min}\propto L^{1/2} , \]
which is just outside the radius of the Broad Line Region
(BLR, Laor \& Draine 1993). Alternatively, the outflow may be
driven by line opacity and be launched from $R_{\rm min}=R_{\rm BLR}$.
However, since $R_{\rm BLR}\propto L^{\sim 1/2}$, as both
observations (Kaspi et al. 2000), and theoretical arguments
(Netzer \& Laor 1993) indicate, then qualitatively similar outflows
could be obtained for dust and line driving.
If, in addition, $\Gamma$ is independent of $L$
(e.g., if the grain size distribution, composition, and dust/gas ratio, 
or the gas ionization level, are the same in all objects) 
we obtain that
\[ v_{\rm max}\propto L^{1/4}, \] 
i.e. a pure luminosity dependence (e.g., Arav, Li, \& Begelman 
1994, eq. 3.3; Scoville \& Norman 1995, eq.~6).
The observed slope $0.62\pm 0.08$ is significantly steeper 
than this simplified model prediction of 0.25. The difference may not 
be significant if the small slope uncertainty is just a coincidence 
due to the small sample size. However, it could also be
due to a failure of one or more of the above assumptions. 
For example, if $R_{\rm min}$ is independent of $L$ then
$v_{\rm max}\propto L^{1/2}$ (e.g., Weymann, Turnshek, \& 
Christiansen 1985).
If the flow is launched at $R_{\rm min}\propto L^{1/2}$,
then the observed slope implies empirically that $\Gamma$ is not
independent of $L$, but rather
$\Gamma\propto L^{0.74}$. 

A correlation analysis using $v_{\rm \tau,max}$ instead of 
$v_{\rm max}$ yields very similar or slightly stronger correlations, 
both here and in the following sections, and for the sake of brevity
we do not report these results. Exclusion of the two lower quality
\IUE-based data points of PG~1011$-$040 and PG~2214+139, also
generally yields similar or somewhat stronger correlations,
and are also not reported here and below.

\subsection{Dependence on other Emission Properties}

BG92 have shown significant correlations among optical emission-line 
parameters,
in particular among the \Hbeta\ width and asymmetry, \MO3,
[\oiii] to \Hbeta\ peak flux ratio, and Fe~II/\Hbeta\ flux ratio
(some of the dominant components of EV1 in BG92). 
These correlations extend to the X-ray (e.g., L97) and to the
UV (e.g., Wills et al. 1999), and may be due to systematic changes in the
BLR and NLR properties with increasing $L/L_{\rm Edd}$.
{\em Are the UV absorption properties of SXWQs related to any of the
BG92 optical emission-line properties?} 

Figure~7 shows that \CEW\ and $v_{\rm max}$ are
significantly correlated with \OEW\ for the SXWQs (in addition
to the low [\oiii] luminosities of SXWQs, \S 3.3), and there
appears to exist an overall upper ``envelope'' of decreasing \CEW\ and 
$v_{\rm max}$ with increasing \OEW, although there is no significant correlation 
for the intermediate and weak absorption AGN. A somewhat weaker 
correlation (Pr=$5\times 10^{-3}$) of \CEW\ and $v_{\rm max}$
was found with \HEW, and none of the 
other emission-line parameters tabulated 
by BG92 shows any significant correlation with either \CEW\ or
$v_{\rm max}$.

The rather strong correlation between \OEW\ and $v_{\rm max}$, together
with the lack of correlation with other EV1 emission-line parameters is
intriguing. 
{\em Is there a direct physical link between $v_{\rm max}$ and}
\OEW? Possible direct links could be a greater destruction of NLR
clouds by higher velocity outflows, or an increasing obscuration of
the NLR due to interactions of a faster outflow with nearby 
dense clouds which increases the scale height of the obscuring gas.
However, there is an indication that the relation between \OEW\ and 
$v_{\rm max}$ may be indirect. There is a very strong correlation of 
\OEW\ with $M_V$ for the SXWQs ($r_S=0.91$, Fig.~5), and there is also a 
very strong correlation of $v_{\rm max}$ with $M_V$ ($r_S=-0.95$, Fig.~6). 
Thus, the somewhat weaker correlation of \OEW\ with $v_{\rm max}$ 
($r_S=-0.90$) may be a secondary effect, as partial correlation analysis
indicates (using eq.~27 in Kendall \& Stuart 1977),
and it does not necessarily imply
a direct physical link. However, the origin of the highly significant
\OEW\ vs. $M_V$ correlation in SXWQs (Pr=$2.8\times 10^{-4}$) remains to 
be understood. AGN in general show a trend of decreasing \OEW\ with 
increasing luminosity (see Fig.~3 in BLW), but with a very broad scatter.
SXWQs populate the lower boundary of the \OEW\ vs. $M_V$ distribution
of AGN, and for some unknown reason this lower boundary decreases 
with increasing luminosity, creating the observed correlation.

Earlier studies suggested that BALQs are restricted to the radio-quiet
quasar population, but recent deep radio surveys are finding
that BALQs occur with about equal frequency in radio-loud and 
radio-quiet quasars 
(Becker et al. 2000; Menou et al. 2001). However,
the UV absorption strength and velocity extent may still
be larger in radio-quiet quasars, in particular when compared to the 
most radio-loud quasars ($R>100$, Becker et al. 2001). Figure~8 shows 
the dependence of \CEW\ and
$v_{\rm max}$ on the radio/optical flux ratio in our sample.
KS tests indicate that the eight radio-loud and 20 radio-quiet AGN 
in our sample 
are consistent with being drawn from the same 
distributions of \CEW\ and $v_{\rm max}$ (Pr=0.49 for both). The 
trend noted in earlier studies may become statistically 
significant if a larger sample is studied.

\subsection{Dependence on Fundamental Parameters}

The black hole mass ($M_{\rm BH}$) in AGN can be estimated
using the \Hbeta\ FWHM ($v_{\rm BLR}$) and the size of the \Hbeta\
emitting region ($R_{\rm BLR}$) obtained from reverberation mapping, or
from the BLR size-luminosity relation (e.g., Kaspi et al. 2000), with the
assumption of virialized motion in the BLR (e.g. Peterson, \& Wandel 2000),
giving $M_{\rm BH}(\Hbeta)= v_{\rm BLR}^2\times R_{\rm BLR}/G$,
or
\[
m_9=0.18v_{3000}^2 L_{46}^{1/2} ,
\]
where  $m_9\equiv M_{\rm BH}({\rm H}\beta)/10^9M_{\odot}$,
 $v_{3000}\equiv v_{\rm BLR}/3000$~km~$^{-1}$, and
$L_{46}$  is the bolometric luminosity in units
of $10^{46}$~erg~s$^{-1}$ (e.g. Laor 1998). Since the Eddington
luminosity in units of $10^{46}$~erg~s$^{-1}$
is $L_{\rm Edd}=12.5m_9$, the above relation implies
\[
 L/L_{\rm Edd}=0.44v_{3000}^{-2} L_{46}^{1/2} .
\]
Such
estimates are potentially significantly uncertain (e.g., Krolik 2001);
however, recent studies have shown that the \Hbeta-based black hole mass
estimate in AGN is correlated with the host bulge
luminosity and with the host bulge stellar velocity dispersion,
as found for
nearby non-active galaxies (Laor 1998; Ferrarese et al. 2001; 
Gebhardt et al. 2000). This indicates that $M_{\rm BH}(\Hbeta)$ 
provides a reasonably accurate estimate of the true $M_{\rm BH}$
(to within a factor of 2--3), and it opens up the possibility of looking
for correlations of various observed properties with the underlying
fundamental parameters, $M_{\rm BH}$ and $L/L_{\rm Edd}$.

The simplified radiation-pressure driven
outflow scenario (\S 3.4) implies that if the outflow is
launched at $R_{\rm min}\simeq R_{\rm BLR}$, then
\[ v_{\rm max}/v_{\rm BLR}\simeq \sqrt{\Gamma L/L_{\rm Edd}}, \]
where we assume $v_{\rm Kep}\simeq v_{\rm BLR}$.
Figure~9 shows there indeed exists an apparently significant correlation
between the observed $v_{\rm max}/v_{\rm BLR}$ and
$L/L_{\rm Edd}$. The value of $L/L_{\rm Edd}$ for each object is
obtained using the \Hbeta\ FWHM from BG92, assuming
$L_{46}=8.3\nu L_{\nu}$(3000~\AA) (Laor 1998), where
$\nu L_{\nu}$(3000~\AA) is
listed in Table~1. The strength of the correlation for the
combined sample of 24 AGN with significant outflow
velocity\footnote{The four objects with upper limits in Fig.~9
show no significant outflow ($v_{\rm max}\le 100$~\kms) and are not 
included in the analysis since the absorption most likely originates
in gas far from the nucleus.} is $r_S$=0.76 (Pr=$1.9\times 10^{-5}$).
The other option mentioned in \S 3.4 that $R_{\rm min}$ is
a constant (independent of $L$) would not imply the correlation
seen in Fig.~9.

 As argued above
(\S\S 3.3, 3.4), the UV absorbers in SXWQs and non-SXWQs are likely
to be different, so we repeated the analysis for the
two populations separately. A marginally significant correlation
is still present for the 10 SXWQs ($r_S=0.72$, Pr=$1.9\times 10^{-2}$), 
and a significant correlation is present for the 14 non-SXWQs
($r_S$=0.85, Pr=$1.6\times 10^{-4}$). The best-fit slopes for the
two populations are very similar; for SXWQs,
$v_{\rm max}/v_{\rm BLR}\propto (L/L_{\rm Edd})^{0.91\pm 0.24}$;
and for non-SXWQs,
$v_{\rm max}/v_{\rm BLR}\propto (L/L_{\rm Edd})^{0.83\pm 0.16}$.
Under the simplified outflow scenario described above these
slopes imply that $\Gamma$ must increase with $L/L_{\rm Edd}$.
Specifically, the above relations imply
$\Gamma\propto (L/L_{\rm Edd})^{0.82}$ for SXWQs,
and $\Gamma\propto (L/L_{\rm Edd})^{0.66}$ for non-SXWQs.
This luminosity dependence of $\Gamma$ is comparable
to $\Gamma\propto (L/L_{\rm Edd})^{0.68}$ suggested by the
$v_{\rm max}$ vs. luminosity relation (\S 3.4),
which is independent of the observed values of $v_{\rm BLR}$.

The non-SXWQs in Fig.~9 fall on average a factor of $\sim 3$ below
the SXWQs. A possible interpretation is that the UV absorbing outflows
in non-SXWQs are launched at $\sim 10$ times larger
distance than in SXWQs. At this larger distance
$v_{\rm Kep}$, and thus the resulting $v_{\rm max}$
(see \S 3.4), are a factor of $\sim 3$
smaller, at a given $L/L_{\rm Edd}$. A possible alternative explanation
is that the launching radius is the same in all objects, but in 
non-SXWQs the outflows become fully ionized before reaching 
$v_{\rm max}$ due to their higher X-ray flux. This alternative cannot
be ruled out, but we note that intermediate $\alpha_{ox}$ AGN
do not have intermediate $v_{\rm max}$, as one may expect given their
apparently intermediate strength ionizing radiation,
but rather the same $v_{\rm max}$ as of
normal $\alpha_{ox}$ AGN of the same $M_V$ (Fig.~6). This alternative
can be directly tested by looking for higher ionization lines (e.g.
\ovi, \neviii) which should display a higher $v_{\rm max}$.

Since $\Gamma\simeq (v_{\rm max}/v_{\rm BLR})^2/(L/L_{\rm Edd})$
(\S 3.4), then Fig.~9 together with our assumed outflow mechanism
suggest that $\Gamma\sim 50-100$
in SXWQs with $L\sim L_{\rm Edd}$.
 Optically thin outflows with
$N_{\rm H}\le 10^{21}$~cm$^{-2}$ driven by dust or resonance-line
opacity can reach $\Gamma\sim 1000$
(e.g., Arav et al. 1994; Scoville \& Norman 1995), but such outflows are too thin
to produce the observed X-ray absorption if their metal abundance
is solar. An optically thick absorber
which absorbs all incident continuum photons can reach at most
$\Gamma\le 1.5\times 10^{24}/N_{\rm H}$, and thus a flow
with $\Gamma\gtsim 50$ would imply  $N_{\rm H}\ltsim 10^{22}$~cm$^{-2}$,
which is optically thin above 1.5~keV (i.e. cannot make an AGN SXW).
Our estimates of $\Gamma$ can
probably be off by an order of magnitude, but if SXWQs with
UV outflows having $\Gamma\gtsim 50$ do exist then the large column density
X-ray absorber of SXWQs should not be part of this outflow, but rather
should form a ``heavy'' non-outflowing ``filter'' for the
the UV absorbing outflow. However, if the X-ray absorber is found to
have the same dynamics of the UV outflow, then the X-ray absorber must
be very ``light'', implying a metallicity well above solar
(i.e., having little H and He which constitute most of the mass, but do not
contribute much to the X-ray opacity).

The $v_{\rm max}/v_{\rm BLR}$ vs. $L/L_{\rm Edd}$ relation is subject
to one major caveat. The two quantities are not determined independently,
but rather both depend on $v_{\rm BLR}$.
Is the observed relation physically
meaningful, or is it just a secondary effect due to the strong 
correlation of $v_{\rm max}$ with
$L$ coupled with the inclusion of $v_{\rm BLR}$ in both the $x$ and $y$
coordinate parameters?  There is no conclusive answer to this question.
However, there
is an indirect indication that the $v_{\rm max}/v_{\rm BLR}$ vs. $L/L_{\rm Edd}$
relation is real. This is provided by the lower panel of Fig.~9 which shows
the relation of $v_{\rm max}/v_{\rm BLR}$ vs. $M_{\rm BH}$. 
The correlation of $v_{\rm max}/v_{\rm BLR}$ with $M_{\rm BH}$ is much
weaker, although the secondary effect mentioned above should
affect this correlation just as well. We carried out additional
tests replacing $v_{\rm BLR}$ with an unrelated optical emission-line
parameter from BG92 (e.g., \OEW, $R$, $z$), which served as a mock
``$v_{\rm BLR}$'', and tested the 
strength of the new mock $v_{\rm max}/$``$v_{\rm BLR}$'' with the new mock 
$L/``L_{\rm Edd}$'' and ``$M_{\rm BH}$''. The results were either a strong
correlation of $v_{\rm max}/$``$v_{\rm BLR}$''
with both $L/``L_{\rm Edd}$'' and ``$M_{\rm BH}$'',
when ``$v_{\rm BLR}$'' had a large range of values in the sample, or 
no significant correlation with both parameters, when ``$v_{\rm BLR}$''
had a small range. Thus, the fact that $v_{\rm max}/v_{\rm BLR}$ 
is significantly correlated with $L/L_{\rm Edd}$, but not with $M_{\rm BH}$,
provides an indirect indication we are detecting a real physical effect.
This result is also consistent with our theoretical expectation that 
$v_{\rm max}/v_{\rm BLR}$ and $M_{\rm BH}$ should not be directly related
(the observed marginally significant correlation could be induced the
correlation of $M_{\rm BH}$ and $L/L_{\rm Edd}$ which has $r_S=-0.66$
in our sample).

\section{Some Further Questions}

\subsection{\em Are there AGN with strong UV absorption
and no X-ray absorption?}
As shown in Fig.~4, there appears to exist a well-defined upper
envelope to the \CEW\ vs. $\alpha_{ox}$ distribution, and thus 
it appears to be impossible to have strong UV absorption without 
strong X-ray absorption. This result is not trivial as one may 
imagine a low-column density absorber ($N_{\rm H}\sim 10^{20}~{\rm cm}^{-2}$) 
which can produce very strong UV absorption (EW$\gg 10$~\AA), but
essentially no
X-ray continuum absorption. {\em Why do such absorbers not exist?}
A possible answer is provided by the model of Murray et al. (1995),
where a strong suppression of the ionizing continuum is essential to
prevent the UV absorbing outflow from becoming fully ionized.
If true, then traces of absorption by highly ionized gas
(e.g., in \ovi, \neviii) may still be detected at higher
outflow velocities in intermediate $\alpha_{ox}$ AGN.

\subsection{\em Are all intermediate-strength UV absorbers the same?}
As described in \S\S 3.3,3.4, the different \MO3\ distributions
of SXWQs and non-SXWQs, in particular at $M_V<-25$, and their different
dependence of
$v_{\rm max}$ on $M_V$, suggest that the
UV absorption in SXWQs and non-SXWQs arises in physically different
types of absorbers. If true, then although both SXWQs and non-SXWQs
can show similar intermediate-strength UV absorption,
higher quality UV spectroscopy could reveal differences in the absorber
physical properties (e.g., covering factors, optical
depths, abundances, depletions). Such differences would provide important
clues about the nature of the UV and X-ray absorbers.

\subsection{\em Are the X-ray and UV absorbers the same?}

The simplest interpretation of the strong link between \CEW\ and
$\alpha_{ox}$ (BLW; Fig.~4) is that the X-ray and the UV absorption
arise in the same absorber. Simple estimates of the UV column
densities in
BALQs assuming optically thin absorption imply column densities of
$N_{\rm H}\sim 10^{19}-10^{20}$~cm$^{-2}$, which are highly
discrepant with the
X-ray deduced column densities of $\sim 10^{23}-10^{24}$~cm$^{-2}$.
Higher quality UV
spectroscopy established the importance of UV line saturation and
partial covering (Kwan 1990; Korista et al. 1992; Arav 1997; Hamann 1998),
and thus increased the implied UV column densities by one
to possibly two orders of magnitude. However, a large discrepancy still
remains in the best-studied BALQ (Arav et al. 2001 vs.
Mathur et al. 2000) and in the best-studied Seyfert absorber
(Kraemer et al. 2001).

The models of Murray et al. (1995) and Murray \& Chiang (1995) suggest
that the X-ray absorbing gas lies at the inner low-velocity part of the
flow and shields the UV outflow from ionizing photons. However, if the
outflow is highly clumped and is driven by radiation pressure on dust
(Scoville \& Norman 1995) or by other mechanisms (e.g., Begelman, de Kool
\& Sikora 1991), then such shielding may not be necessary. Thus, an
observational determination of the relation between the UV and X-ray
absorbers may allow discrimination between different acceleration
models.

If the X-ray and UV absorbers are the same then, apart from a similar
column density, they should also share the same kinematics. The available
UV spectroscopy allows an accurate determination of the outflow kinematics
but typically only weak constraints on the column density due to
partial-covering and line-saturation effects,
while the opposite is true for the available low-resolution
X-ray spectroscopy. This problem can be alleviated, at least in part,
in two ways. On the UV side, more accurate column densities can be
obtained through high-quality spectroscopy of low-luminosity SXWQs,
where the low outflow velocities allow one to resolve various
absorption doublets, and thus better constrain the UV absorbing column
density (e.g., PG~1351+64, Zheng et al. 2001).
On the X-ray side, more accurate kinematic constraints can be
obtained through X-ray grating spectroscopy of relatively X-ray bright
SXWQs, where the outflow velocities should
 be fairly easy to resolve if they are similar to the observed UV
 velocities, even if the S/N is not very high (e.g., PG~2112+059,
Gallagher et al. 2001). The intermediate X-ray and UV absorption
AGN (Figs.\ 2, 4) could be the optimal objects for applying the
absorber kinematics ``identity test''; they are not as heavily
absorbed in the X-rays as SXWQs are, their UV absorption is
generally narrow enough to allow doublet resolution, but the
outflow velocity is large enough to be resolved with X-ray
grating spectroscopy. Such tests are now being done for nearby
bright Seyfert galaxies with relatively weak UV and X-ray
absorption (e.g., Kaspi et al. 2001), they may be possible
for the intermediate objects (with a fairly large investment of time),
and they may be extended to ``proper'' SXWQs in the future when
larger collecting area X-ray telescopes become available.

\subsection{\em What controls the outflowing column density?}
The results presented here suggest that $v_{\rm max}$
is largely set by the luminosity, as some radiation-pressure scenarios
suggest. In a source with a given $L/L_{\rm Edd}$,
radiation pressure can just overcome gravity and start
a radially accelerating optically thick outflow for
$N_{\rm H}=1.5\times 10^{24}f L/L_{\rm Edd}~{\rm cm}^{-2}$,
where $f$ is the fraction of flux absorbed by the outflow
from the total incident flux.
For example, objects with
$f L/L_{\rm Edd}\ltsim 0.01$ should not be able to drive
an outflow with $N_{\rm H}>1.5\times 10^{22}~{\rm cm}^{-2}$
and thus cannot produce a hard X-ray absorbing outflow
($\tau>1$ at 2~keV).
Combined X-ray and UV spectroscopy may allow us to measure
the $N_{\rm H}$ of the outflow and test if it is indeed set by
$f L/L_{\rm Edd}$.

\subsection{\em Why are SXWQs with luminous [\oiii] rare?}

As discussed in BLW and in \S 3.3, SXWQs are $\sim 10$ times more common
in AGN with lower [\oiii] luminosity (\MO3$>-27$) compared to AGN with
luminous [\oiii] (\MO3$<-27$), and the difference in \MO3\ distribution
is most apparent for luminous AGN ($M_V<-25$). Since
the measured luminosity of [\oiii] is most likely not inclination dependent,
this implies the existence of ``luminous [\oiii]'' AGN
which for some reason are only
rarely SXW. Do these AGN just have a $\sim 10$ times lower covering factor
of high-column density clouds ($N_{\rm H}\ge 10^{22}~{\rm cm}^{-2}$)
capable of X-ray absorption (\S 3.1)?  Most likely not since
models of the BLR indicate that $N_{\rm H}> 10^{22}~{\rm cm}^{-2}$ clouds
are required to produce the low-ionization lines (e.g., Netzer 1990),
which are present at about comparable strengths in both low and high
[\oiii] luminosity AGN.
A possible alternative explanation, which also physically relates the X-ray
absorption to the weakness of [\oiii], is that luminous [\oiii] AGN do
not lack high-column density clouds, but rather lack high-column density
clouds with a low dust/gas ratio. For some reason, some AGN may have in
addition to the normal BLR gas component, another high-column density
component with a low dust/gas
ratio\footnote{The composite spectrum of Brotherton et al. 2001
suggests some UV reddening in BALQs, and thus a dust column of
$\sim 10^{21}$~cm$^{-2}$, which is $\sim 10^{-2}-10^{-3}$ of the
column expected based on the X-ray absorption and a normal
Galactic dust/gas ratio.}
covering a large solid angle (a strong wind where the dust is
mostly destroyed?), through which an AGN would appear SXW,
and this component may also filter much of the ionizing radiation from
reaching the NLR, thus explaining why the [\oiii] luminosity is low. 
If true, then this large solid angle NLR obscuring material may
also have other observational consequences, such as on the IR spectral
energy distribution and on the optical polarization.

\section{Conclusions}
We describe the results of a study of the \civ\ absorption properties of 
AGN extending from the Seyfert level ($M_V\simeq -21$) to the luminous
quasar level
($M_V\simeq -27$). The study is based on spectra of an incomplete subset 
of 56 AGN from the 87 BG92 AGN, mostly extracted from the \HST\ archives. 
The main results are the following:

\begin{enumerate}
\item About 10\% of the objects show strong (EW$>$10~\AA) \civ\ absorption,
$\sim 20$\% show intermediate-strength (1~\AA--10~\AA) absorption,
$\sim 20$\% show weak (0.1~\AA--1~\AA) absorption, and
$\sim 50$\% show no ($<$0.1~\AA) absorption.

\item SXWQs are $\sim10$ times more common at \MO3$>-27$ compared
to \MO3$<-27$, but the frequency of UV absorption in objects without
strong X-ray absorption is independent of \MO3. This indicates that 
UV absorption without strong soft X-ray absorption may
be purely an inclination effect, but apparently higher column
UV + soft X-ray absorption (yet without much optical extinction), as also
seen in BALQs, occurs in a physical
component present only in certain types of AGN.
 
\item SXWQs have a higher \CEW\ and $v_{\rm max}$ at a given $M_V$
than UV-only absorbed AGN, and both parameters in SXWQs  are 
strongly correlated with $M_V$ ($r_S=-0.93$ and $-0.95$). Thus, luminous SXWQs
are BALQs, and lower luminosity SXWQs have ``mini-BALs'' or
``associated absorbers''. The observed dependence
$v_{\rm max}\propto L^{0.62\pm 0.08}$ is steeper than expected for
constant-$\Gamma$, radiation-pressure driven outflows
launched at $R\sim R_{\rm BLR}$. This may indicate that $\Gamma$
increases with luminosity.

\item There is an indication that the relative outflow
velocity, $v_{\rm max}/v_{\rm BLR}$, increases with $L/L_{\rm Edd}$
for all AGN, and that the wind in non-SXWQs is launched at $\sim 10$
times the launching radius in SXWQs.

\end{enumerate}

The results presented here are based on UV spectra of 56 of the BG92 
AGN, a few of which have only low-quality spectra. To establish
the strength and statistical significance of these results it is
important to obtain high-quality UV coverage of the complete BG92 
sample of 87 AGN. This survey, together with the BLW study, will 
establish complete UV + soft X-ray coverage of the sample. This can 
then be followed by detailed UV and X-ray spectroscopy of all the 
``interesting'' AGN, which may lead to solutions of some of the
above questions. In particular, the class of ``intermediate'' UV and X-ray
absorption AGN (seven identified here, Fig.~2) may be
especially suitable for studying the relation of the UV and X-ray absorbers.

This first complete UV survey of low-$z$ AGN, together with
ground-based surveys of
high-redshift AGN samples, will also allow one to explore if there is 
significant evolution in AGN absorption properties with $z$. A complete survey
is a resource-intensive approach but, unlike the study of individual 
AGN, will allow one to draw conclusions about the general optically
selected AGN population.
In addition, it is important to define a significantly larger complete 
sample of SXWQs (e.g., Risaliti et al. 2001) to test the strength and 
significance of the relations described above.

We thank Todd Boroson for providing us with accurate redshifts for
all objects in the BG92 sample, and Bev Wills and Sarah Gallagher
for their helpful comments. We also thank the referee for a very
detailed and helpful report.
This research was supported by grant \# 209/00-11.1 of
the Israel Science Foundation to A.L. and NASA LTSA grant NAG5-8107
to W.N.B.

\appendix
\begin{center}A. NOTES ON INDIVIDUAL OBJECTS\end{center}

Relevant notes on some of the objects appear in BLW (\S\S 5,6).
Here we provide additional notes, mostly on objects with weak
\civ\ absorption.

0003+158-- BLW note \CEW=0.8~\AA\ for this object, which refers to
the $z=0.366$ metal absorber noted by Jannuzi et al. (1998, hereafter J98). 
Since the associated
\Lya\ line is unresolved (FWHM$<250$~\kms), the velocity shift of the
system is very large ($v=-17,500$~\kms), and the higher order Lyman series
lines indicate a (partially) optically thin absorber, this is
likely to be an intervening system and we adopt here \CEW=0~\AA.
However, we cannot rule out an intrinsic narrow high-velocity system, 
as suggested in some high-redshift quasars (e.g., Jannuzi et al. 1996;
Hamann, Barlow, \& Junkkarinen 1997; Richards et al. 1999),
though the absorption in the more secure such cases is significantly
broader than 250~\kms.
 
0007+106-- \HST\ did not observe \civ. There may be very weak 
\nv\ absorption (EW$\sim 0.2$~\AA), as also suspected by Crenshaw 
et al. 1999 (\S B1 there). The \Lya\ EW used here is likely an 
overestimate of the \CEW. No significant X-ray absorption is
seen with \ROSAT\ (Wang et al. 1996).

0050+124-- Weak absorption in \Lya, \nv, and \civ\ was
noted by L97 and Crenshaw et al. (1999).
A closer inspection reveals also the 
\siIV$\lambda\lambda$1393.8,1402.8 doublet in absorption
at the same velocity shift of $-1850$~\kms. Small excess absorption
may be present in soft X-rays (Boller, Brandt \& Fink 1996).

0844+349-- \HST\ did not observe \civ. Wang et al. (2000) 
noted the $v\sim 0$~\kms\ absorption system in \Lya. The \nv\ 
absorption is weak (EW$\sim 0.1$~\AA), but both doublet 
components appear to
be present. Wang et al. suggest in addition that the absorption at 
observed-frame 1260.4~\AA\ (rest-frame $-7770$~\kms) 
is too strong and too broad to be just due to Galactic 
\siII$\lambda 1260.4$ absorption and conclude there is
intrinsic \Lya\ absorption in this feature as well. However, the
Galactic \cii$\lambda 1335$ absorption has a very similar
profile (see Fig.~3), which suggests the 1260.4~\AA\ feature is
pure Galactic absorption. There is also evidence for a small 
excess ``cold'' X-ray absorbing column density in an \ASCA\ observation
(George et al. 2000, hereafter G00). Note that this object shows 
large amplitude
changes in $\alpha_{ox}$ (Gallagher et al. 2001, Fig.~9 there).

0923+201-- \HST\ did not observe \civ, and only part of the blue
wing of \nv\ was observed. However, the ``valley'' between \nv\
and \Lya\ is much deeper than normally seen in AGN spectra, where
the blue wing of \nv\ generally blends into \Lya\ with only a
slight drop in flux density (see all the other profiles in Figs. 1-3).
The bottom of this dip at $v\sim -2500$~\kms\ matches
quite well the \Lya\ absorption system at $-2500$~\kms. The
absorption in both \Lya\ and \nv\ appears to include a broad,
shallow component ($-4000\ltsim v\ltsim -2000$~\kms) on top of a
narrow unresolved core.

0947+396-- BLW suggest weak \civ\ absorption (EW$\sim 0.2$~\AA)
based on the line-peak asymmetry. Since there is no clear absorption
dip in the line profile we adopt a more conservative estimate
of EW=0~\AA. \ROSAT\ and \SAX\ spectra do not find significant
absorption (L97; Mineo et al. 2000, hereafter M00).

0953+414-- J98 did not note intrinsic \civ\ absorption in this
object, but Ganguly et al. (2001) did. The \civ\ absorption is 
only marginally significant, though both doublet components seem 
to be present. The velocity shift of $-1380$~\kms\ found by 
Ganguly et al. is based on $z=0.239$. The revised value of
$z=0.23405$ based on the peak of [\oiii]$\lambda 5007$ 
(see Table~1) gives a velocity shift of only
100~\kms. \ROSAT\ and \ASCA\ spectra (L97; G00) suggest there 
may be a small intrinsic X-ray 
absorbing column density.

1011$-$040-- BLW give \CEW$<$1.2~\AA\ based on the available \IUE\
spectrum. However, the marginally significant absorption
centered at $v= -200$~\kms\ in \civ\ appears to be present in
\nv\ and \Lya\ as well. We therefore adopt \CEW$=1$~\AA\ for this 
object, though a higher quality spectrum is obviously required
to verify this result.  

1049$-$005--  The \civ\ absorption was noted by J98 and by 
Ganguly et al. (2001). The absorption is very weak, but both doublet
components are clearly present.

1100+772-- The \civ\ absorption was noted by J98 and by 
Ganguly et al. (2001). The absorption is rather weak but clearly present 
in \civ, \Lya, and \ovi. \ROSAT\ and \ASCA\ spectra do not reveal any
intrinsic absorption (Wang, Brinkmann, \& Bergeron 1996; Sambruna et al. 
1999).

1115+407-- The \civ\ absorption is weak, but both doublet components
are clearly present. The \nv\ doublet is also clearly present.
\ROSAT\ and {\it BeppoSAX} do not reveal significant intrinsic 
absorption (L97; M00).

1116+215--  BLW suggest very weak \civ\ absorption (EW$\sim 0.1$~\AA)
based on a feature at $v= -9500$~\kms. This feature is below
the detection threshold of J98, and the second doublet component is
not seen. If real, this system is likely to be associated with the
intervening \Lya\ system at $v= -9500$~\kms (related to
an intervening galaxy; Tripp, Lu \& Savage 1998), and we therefore adopt
here \CEW=0~\AA. A very high S/N \ROSAT\ spectrum puts a strong constraint
on any cold absorber (L97), and an \ASCA\ spectrum 
suggests a weak warm absorber (G00).

1202+281-- BLW give \CEW=0.4~\AA. This is based on the \civ\ line-peak 
asymmetry. However, inspection of the \Hbeta\ profile in BG92
reveals the same asymmetry. Since \Hbeta\ absorption is extremely rare,
we conclude that the emission-line asymmetry is most likely not due to
absorption, and adopt here \CEW=0~\AA. A rather high S/N \ROSAT\ spectrum
does not reveal any evidence for absorption (L97)

1211+143-- BLW give \CEW=0.5~\AA. This is based on marginally significant
features at $v\sim -8000$~\kms. The two apparent narrow \civ\ components 
are somewhat too close (400~\kms\ apart, instead of 500~\kms) and are
displaced by $\sim 200$~\kms\ from the nearest \Lya\ absorber. We conclude
that the identification of this system is not secure and adopt \CEW=0~\AA.
High S/N \ROSAT\ and \ASCA\ spectra do not reveal any 
intrinsic absorption (Fiore et al. 1994; G00).

1322+659-- BLW give \CEW=0.2~\AA, based on a narrow absorption
feature at $v=120$~\kms. A similarly shaped absorption feature occurs
in \Lya, but it is centered at $v=-20$~\kms. The apparent \civ\ absorption
is centered at 1808.2~\AA, and it may be due to Galactic 
\siII$~\lambda 1808.0$ since the other resonance \siII\ lines at
1190.4~\AA, 1193.3~\AA, 1260.4~\AA, 1304.4~\AA, and 1526.7~\AA\ are 
unusually strong. We therefore conservatively adopt 
\CEW=0~\AA. A \ROSAT\ spectrum suggests a low-column density intrinsic neutral 
absorber, and an \ASCA\ spectrum suggests a weak warm absorber (L97; G00).
 
1352+183-- BLW give \CEW=0.5~\AA\ based on a narrow absorption
feature at $v=-9900$~\kms. However, since the second \civ\
doublet component does not appear to be present, and the nearest
possible \Lya\ absorption is at $v=-10200$~\kms, we conclude that this
feature is most likely not due to \civ\ and adopt \CEW=0~\AA.
No absorption is seen with \ROSAT\ and \SAX\
(L97; M00).

1402+261-- Although \civ\ absorption is clearly seen at
$v=-4600$~\kms, two earlier \HST\ observations in 1993 (PI 
Tytler) and in 1994 (Turnshek et al. 1997) do not show significant
absorption in \civ, providing strong evidence that the absorption
is variable. No absorption is seen with \ROSAT\ and \SAX\
(L97; M00).

1427+480-- BLW give \CEW=0.03~\AA\ based on two very weak features
at $v=-100$~\kms\ and $v=400$~\kms, and a weak \Lya\ absorption
feature at $v=0$~\kms. Since the amplitudes of these features are 
comparable
to the noise we adopt here \CEW=0~\AA. No absorption is seen with \ROSAT\ 
(L97).

1512+370-- BLW give \CEW=0.2~\AA\ based on two apparently significant
features at $v=250$~\kms\ and $v=800$~\kms. However, since there is no
clear corresponding \Lya\ feature we adopt \CEW=0~\AA\ here. J98 and
Ganguly et al. (2001) also do not identify these features as \civ\ 
absorption. No absorption is seen with \ROSAT\ and \SAX\
(L97; M00).

1543+489-- BLW give \CEW=0.4~\AA\ based on an apparently very broad
and very shallow absorption dip at the \civ\ line peak ($v=-1000$~\kms\ 
to $-3000$~\kms). However, since we cannot rule out a peculiar line
profile in this rather extreme EV1 object (BG92), we adopt \CEW=0~\AA\
here.

2214+139--  BLW give \CEW$<1.2$~\AA\ based on the available \IUE\
spectrum. However, the peak of \nv\ appears to be
redshifted by $\sim 1000$~\kms, which is highly unusual and
may imply absorption of the blue wing of \nv. Some absorption
may be present in \civ\ at $v=-1000$~\kms, 
though \Lya\ shows no evidence for absorption.  
We tentatively estimate a marginally significant 
\CEW=1.1~\AA\ and use a different symbol for this object 
throughout the paper. 

2251+113-- BLW give \CEW=0.8~\AA\ based on a very deep resolved 
\civ\ doublet
at $v=0$~\kms. Here we identify additional absorption extending up to
$v=-5000$~\kms, which increases \CEW\ to 3.5~\AA\ (see also the comment
in \S 6.4 of Ganguly et al. 2001).
The velocity shift of $610$~\kms\ found by 
Ganguly et al. is based on $z=0.323$. The revised value of
$z=0.32553$ (see Table~1) gives a velocity shift of 0~\kms.

\appendix
\begin{center}B. OBJECTS WITH \CEW$<0.1$~\AA \end{center}

For the sake of completeness we list the 28 objects where we
find intrinsic \CEW$<0.1$~\AA. This limit is based on \civ, \Lya, and \nv.
In objects where these three lines are not all available we list
in parentheses the line which was available and used to
constrain the UV absorption. The ``no UV absorption'' sample includes
the following objects:
PG~0003+158,
PG~0003+199, PG~0026+129, PG~0052+251 (\civ), PG~0947+396,
PG~1103$-$006, PG~1116+214, PG~1121+422 (\civ), PG~1202+281,
PG~1211+143, PG~1216+069, PG~1226+023, PG~1259+593,
PG~1302$-$102, PG~1307+085 (\civ), PG~1322+659, PG~1352+183,
PG~1415+451, PG~1416$-$129, PG~1427+480, PG~1440+356,
PG~1444+407, PG~1512+370, PG~1534+580 (\Lya), PG~1543+489,
PG~1545+210, PG~1612+261 (\civ), PG~1626+554.

\clearpage

\begin{table}
\caption{\civ\ Absorption Parameters for BG92 Quasar Sample}
\begin{tabular}{lccccccrr}
\tableline \tableline
Object & $z^a$ & $M_V^b$ & $\nu L_{\nu}^c$ & $\alpha_{ox}$ & \OEW\ & \CEW\ & $v_{\rm max}$ &
$v_{\rm \tau,max}$ \\
name & &  & & & (~\AA) & (~\AA) & (\kms) & (\kms) \\
\tableline
0007+106$^d$ & 0.0894 & $-$23.85 & 44.77 & $-$1.43 & 42 & 0.8  &  360  &   360 \\
0043+039$^e$ & 0.3859 & $-$26.16& 45.49 &$<-$2.00 &  1 & 22.3 & 19000 & 11000 \\
0050+124 & 0.0609 & $-$23.77& 44.58 & $-$1.56 & 22 & 0.4  &  1850 &  1850 \\
0844+349$^d$ & 0.0644 & $-$23.31& 44.46 & $-$1.54 &  8 & 0.6  &  $-$100 &  $-$100 \\
0923+201$^d$ & 0.1929 & $-$24.56 & 45.14& $-$1.57 &  7 & 0.8  &  2700 &  2700 \\
0953+414 & 0.2340 & $-$25.65& 45.47 & $-$1.50 & 18 & 0.1 &   100 &   100 \\
1001+054 & 0.1610 & $-$24.07& 44.81 & $-$2.13 &  7 & 11.8 & 10000 &  6150 \\
1004+130$^f$ & 0.2404 & $-$25.97& 45.47 &$<-$2.01 &  6 & 16.6 & 12000 &  6800 \\
1011$-$040$^f$ & 0.0584 & $-$22.70& 44.22 & $-$2.01 & 15 &  1.0: &  3000: &  2000: \\
1049$-$005 & 0.3596 & $-$25.93& 45.54 & $-$1.56 & 55 &  0.3 &  4050 &  4050 \\
1100+772 & 0.3115 & $-$25.86 & 45.54& $-$1.39 & 41 &  0.4 &   750 &   750 \\
1114+445 & 0.1438 & $-$24.01& 44.67 & $-$1.62 & 17 &  4.0 &   800 &   250 \\
1115+407 & 0.1542 & $-$23.74 & 44.55& $-$1.45 &  6 &  0.3 &  $-$200 &  $-$200 \\
1126$-041^f$ & 0.0601 & $-$23.00& 44.34 & $-$2.13 & 19 &  5.8 &  3300 &  2000 \\
1309+355 & 0.1825 & $-$24.76 & 44.92& $-$1.71 & 19 &  2.2 &  1450 &   900 \\
1351+640 & 0.0880 & $-$24.08& 44.78 & $-$1.78 & 31 &  5.3 &  2300 &   900 \\
1402+261$^e$ & 0.1643 & $-$24.48& 45.04 & $-$1.58 &  1 &  0.7 &  4600 &  4600 \\
1404+226$^e$ & 0.0978 & $-$22.93& 44.13 & $-$1.55 &  7 &  1.5 &  2400 &  1800 \\
1411+442 & 0.0898 & $-$23.54& 44.52 & $-$2.03 & 15 & 10.3 &  4100 &  2000 \\
1425+267 & 0.3635 & $-$26.18& 45.22 & $-$1.63 & 36 &  2.1 &  1600 &   800 \\
1535+547$^{d,g}$ & 0.0388 & $-$22.15& 43.82 & $<-$2.17 & 16 &  3.9 &  1400 &   400 \\
1700+518 & 0.2892 & $-$26.44 & 45.51&$<-$2.29 &  0 &  94  & 31000 & 12000 \\
1704+608 & 0.3721 & $-$26.38& 45.61 & $-$1.62 & 27 &  2.4 &  2500 &  1750 \\
2112+059 & 0.4597 & $-$27.26 & 46.13& $-$2.11 &  0 & 26.7 & 24000 & 18400 \\
2130+099 & 0.0631 & $-$23.23& 44.62 & $-$1.47 & 20 &  0.6 &  1550 &  1550 \\
2214+139$^f$ & 0.0657 & $-$23.39& 44.46 & $-$2.02 &  9 &  1.1: &  2000: &  1000: \\
2251+113 & 0.3255 & $-$26.24& 45.63 & $-$1.86 & 19 &  3.5 &  5000 &  2950 \\
2308+098 & 0.4336 & $-$26.24& 45.62 & $-$1.35 & 17 &  0.2 &     0 &     0 \\
\tableline
\normalsize
\end{tabular}\\
$^a$ Kindly provided by T. A. Boroson.\\
$^b$ Taken from BG92.\\
$^c$ In units of log ergs~s$^{-1}$, based on $f_{\nu}$ at rest frame
3000~\AA\ from Neugebauer et al. (1987), calculated for
$H_0=80$~km~s$^{-1}$~Mpc$^{-1}$, $\Omega_0=1.0.$\\
$^d$ Based on \Lya.\\
$^e$ No Galactic absorption lines were available for calibration.
The velocity scale may be off by $\sim 300$~\kms.\\
$^f$ Based on a low spectral resolution ($\sim 1000$~\kms)
\IUE\ spectrum.\\
$^g$ The upper limit on $\alpha_{ox}$ is from Gallagher et al. (2001).\\
: Marginally significant results.\\
Note: the following five objects have a non zero balnicity index
(Weymann et al. 1991):
PG~0043+039, 1210~km~s$^{-1}$;
PG~1001+054, 684~km~s$^{-1}$;
PG~1004+130, 417~km~s$^{-1}$;
PG~1700+518, 13140~km~s$^{-1}$;
PG~2112+059, 1520~km~s$^{-1}$.
\end{table}

\begin{table}
\caption{Estimated Statistics of Obscuring Material Covering Factors}
\begin{tabular}{lccc}
\tableline \tableline
 & \CEW & \MO3$<-27$ & \MO3$>-27$ \\
\tableline
Clear & $<0.1$~\AA &  $\sim 57$\%\ (17/30) & $\sim 40$\%\ (11/26)\\
 & & \\
Weak absorption & 0.1--1~\AA & $\sim 20$\%\ (6/30) & \\
 & & & $\sim 25$\%\ (6/26)\\
Intermediate absorption & 1--10~\AA& $\sim 20$\%\ (6/30) & \\
 & & \\
Strong absorption & $>10$~\AA & $\sim 3$\%\ (1/30) & $\sim 35$\%\ (9/26) \\
\tableline
\normalsize
\end{tabular}\\

\end{table}

\clearpage
\begin{figure}
\plotone{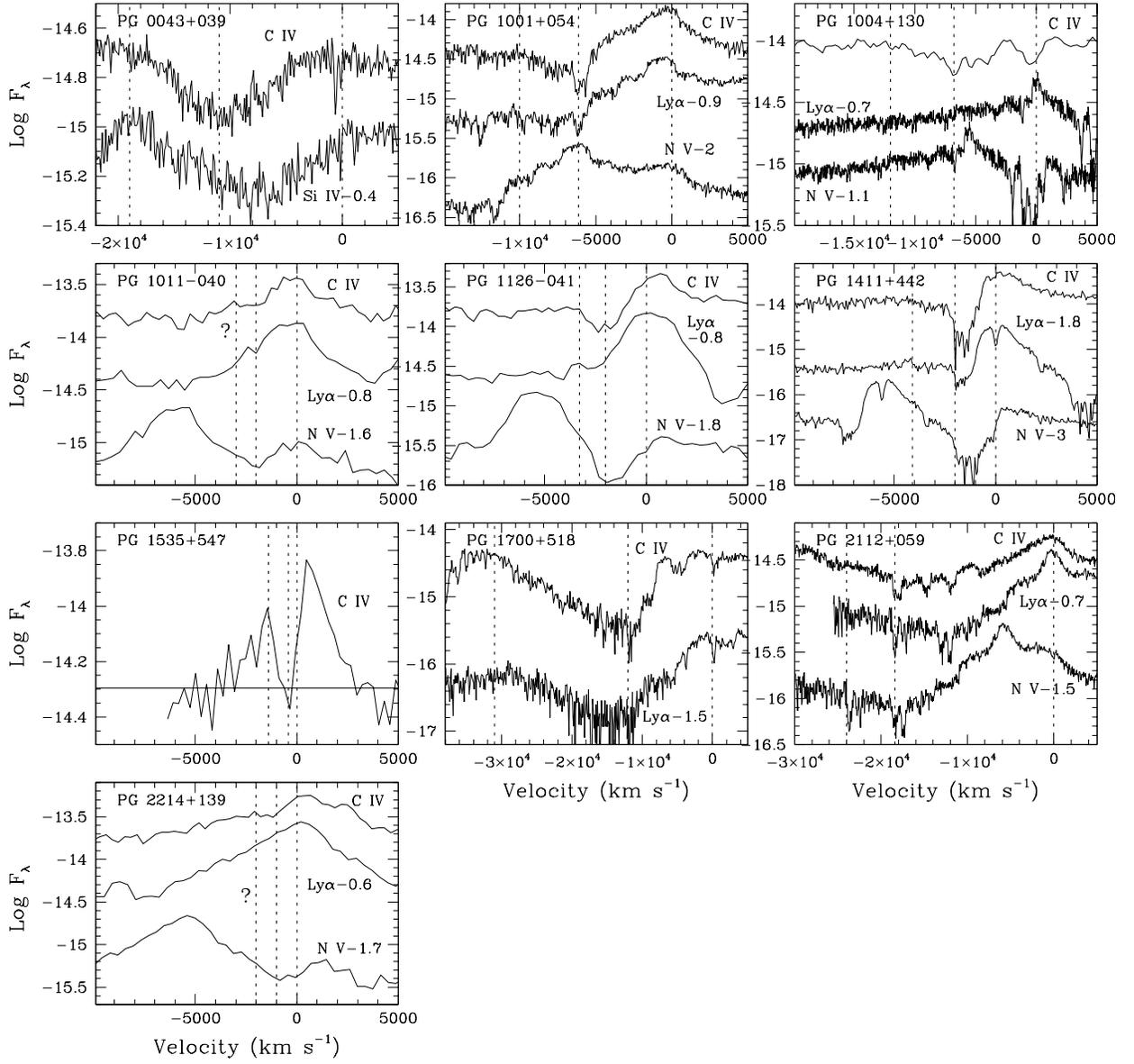}
\caption{The \civ, \Lya, and \nv\ absorption-line profiles for the 
SXWQs (not all lines are available for all objects). The flux density
is in units of observed ergs~s$^{-1}$~\AA$^{-1}$~cm$^{-2}$.
The three
vertical dashed lines in each panel indicate $v=0$~\kms, defined with
respect to the shorter wavelength doublet component, the velocity
of maximum absorption, $v_{\rm \tau,max}$ and the maximum velocity 
of absorption, $v_{\rm max}$.
The \Lya\ and \nv\ profiles are shifted downward for clarity by
the amount indicated next to the line label.
Note that the significance of the absorption in PG~1011$-$040 and
PG~2214+139 is only marginal.}
\end{figure}

\begin{figure}
\plotone{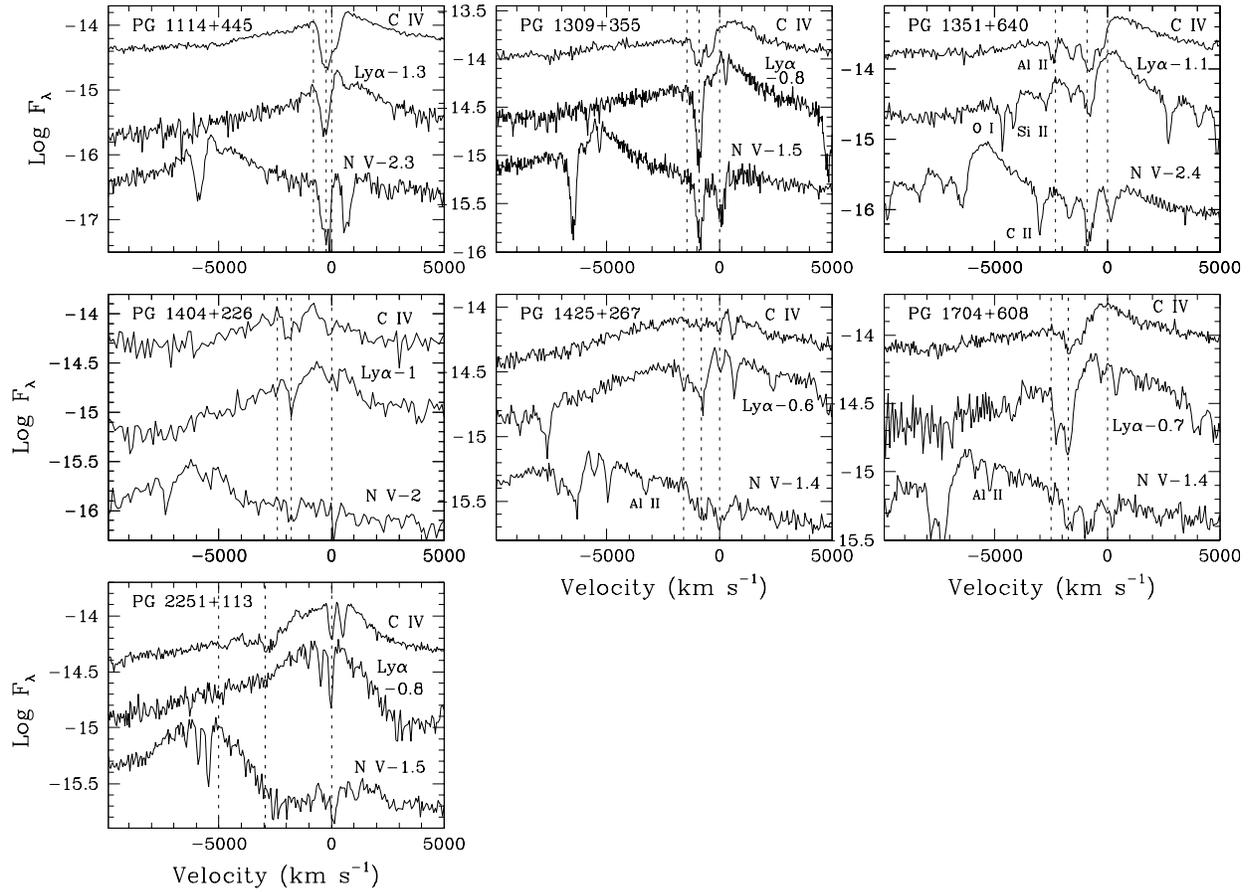}
\caption{As in Figure 1 for the objects with intermediate-strength
\civ\ absorption (1~\AA$<$EW$<$10~\AA) that are not SXWQs. 
Note that since the absorption lines are generally narrow, 
it is possible to resolve the longer wavelength
doublet component of \civ\ at 500~\kms, and that of \nv\
at 960~\kms. Narrow absorption features due to Galactic absorption
are designated.}
\end{figure}

\begin{figure}
\plotone{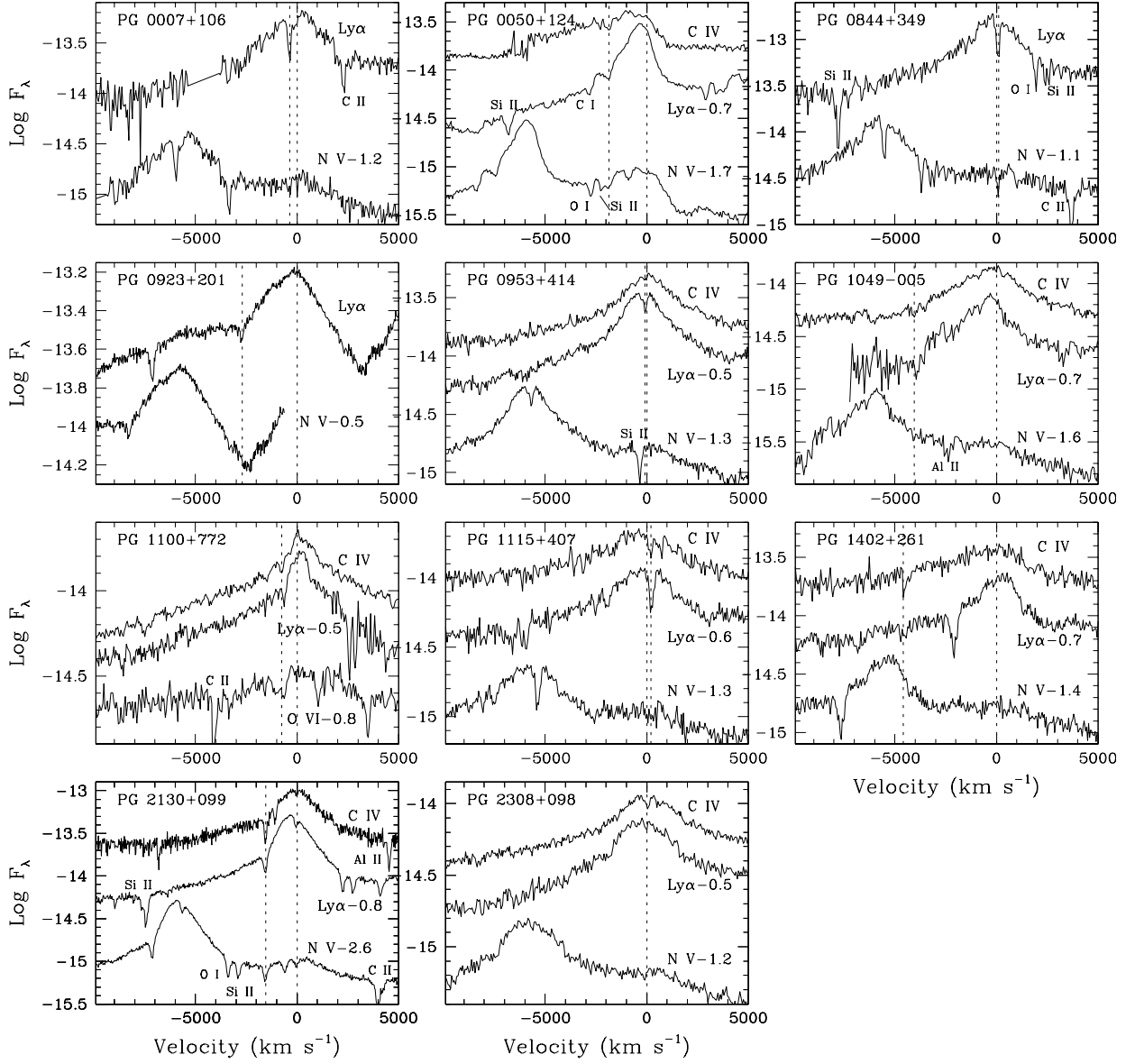}
\caption{As in Figures 1 and 2 for the objects with weak
\civ\ absorption (EW$<$1~\AA). Since the absorption features are
not resolved, only $v_{\rm \tau,max}$ is indicated.}
\end{figure}

\clearpage
\begin{figure}
\plotone{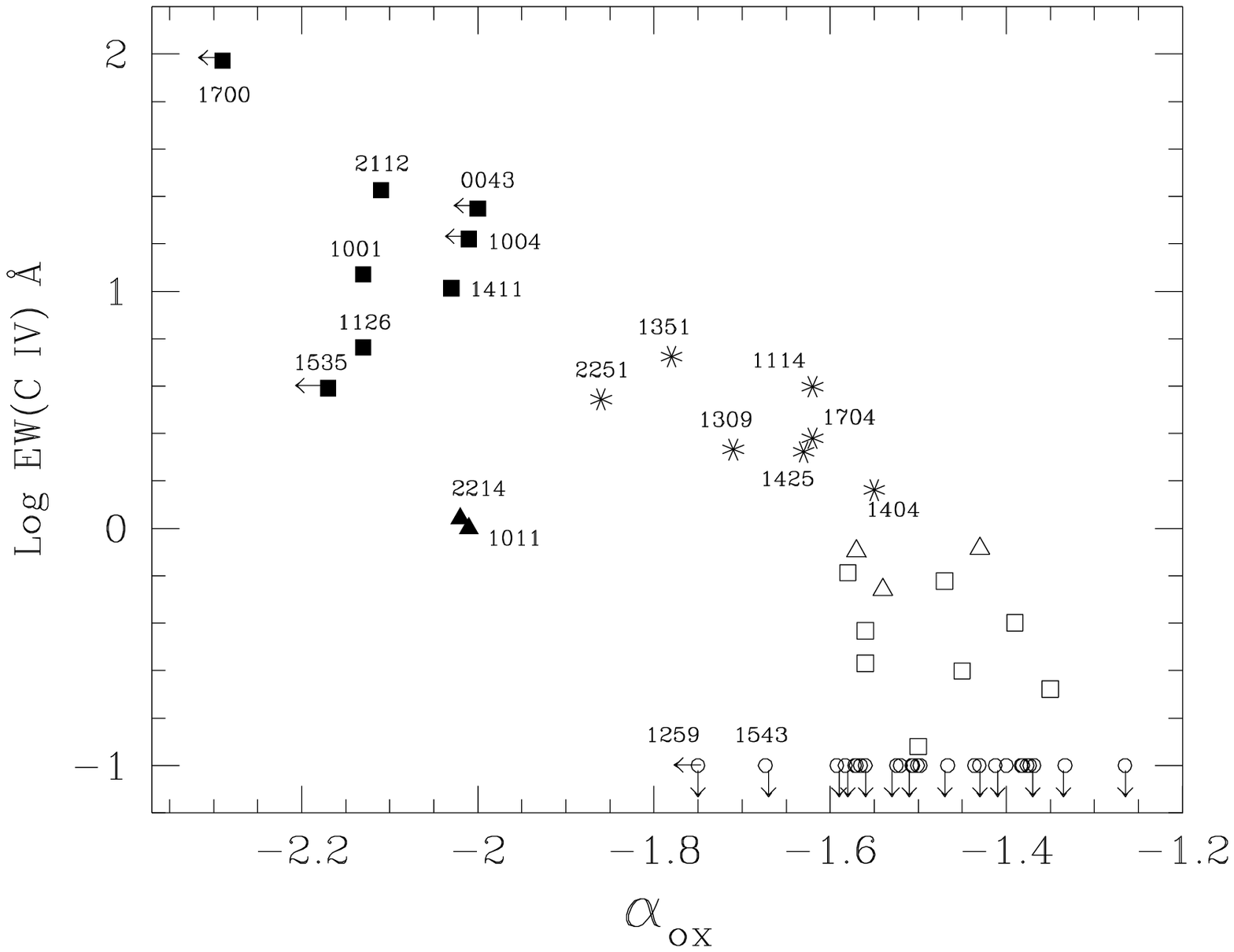}
\caption{The absorption \CEW\ vs. $\alpha_{ox}$ relation for the BG92 AGN.
Filled squares are SXWQs, stars are non-SXWQs with intermediate 
\civ\ absorption (1~\AA$<$EW$<$10~\AA), open squares are AGN with
weak \civ\ absorption (EW$<$1~\AA), and open triangles have weak
\Lya\ absorption. Open circles at log~\CEW$=-1$ indicate objects
with no detectable absorption. Filled triangles mark the two SXWQs 
where the absorption is only marginally significant. 
Some objects are designated by the right ascension part of their name. 
Note that AGN with intermediate absorption have a steeper $\alpha_{ox}$
than AGN with weak absorption, and all AGN with strong
absorption (EW$>$10~\AA) are SXW.}
\end{figure}

\clearpage
\begin{figure}
\plotone{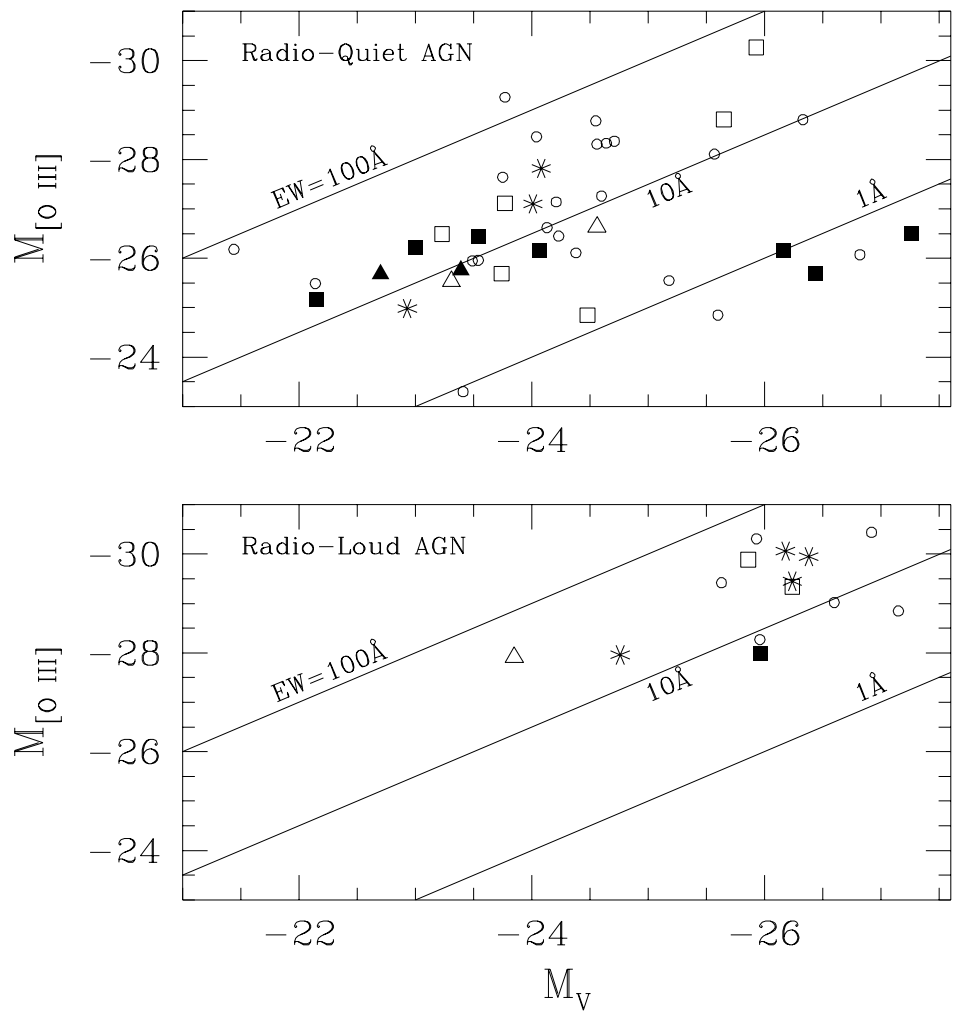}
\caption{The \MO3\ vs.
$M_V$ relation for the radio-quiet and radio-loud BG92 AGN. 
Symbols are as in Fig.~4, and open circles
mark the 28 objects with no \civ\ absorption. The diagonal
solid lines indicate constant [\oiii] EW. The distribution of
AGN with intermediate/weak UV absorption is similar to 
AGN with no absorption, but SXWQs have preferentially low
[\oiii] luminosity.}
\end{figure}

\clearpage
\begin{figure}
\plotone{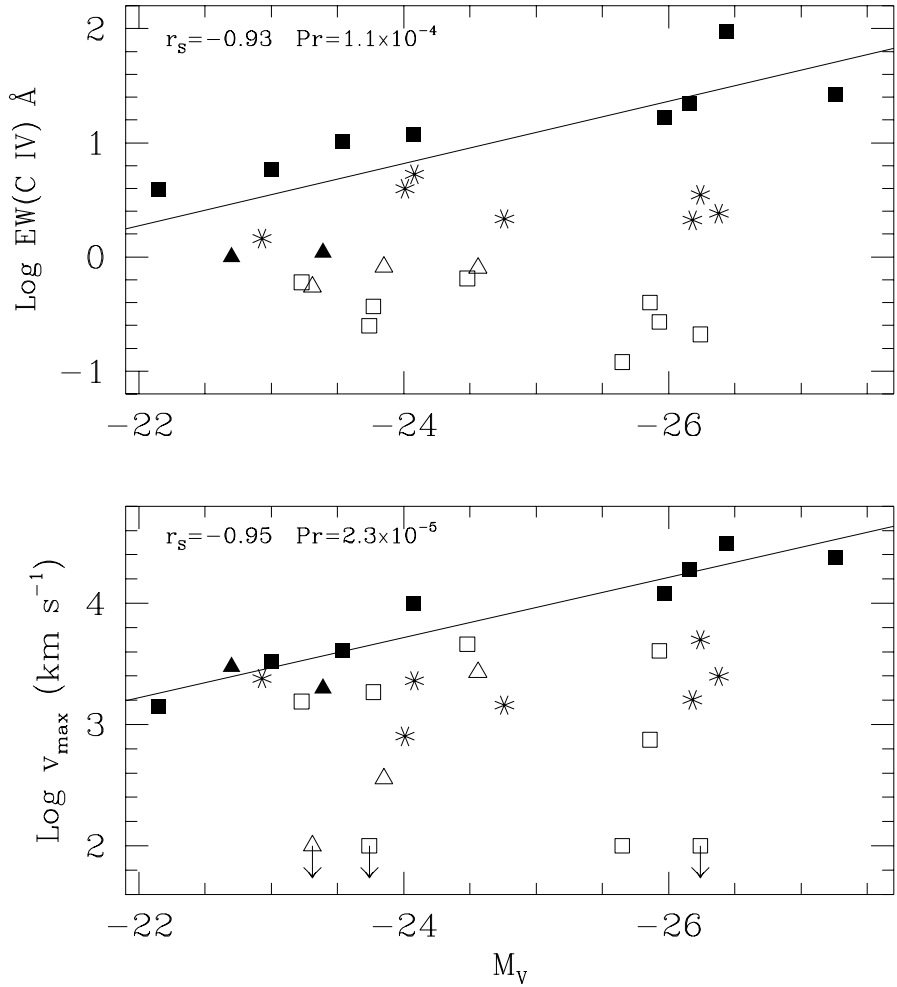}
\caption{The luminosity dependence of \CEW\ and of $v_{\rm max}$. 
SXWQs have higher \CEW\ and $v_{\rm max}$ than other absorbed
quasars at any given luminosity, and both parameters of the SXWQs are 
significantly correlated 
with luminosity. The Spearman correlation coefficient for
the SXWQs ($r_{\rm S}$) and its significance level (Pr) 
are indicated in each panel.
The solid line in each panel is a least-squares fit to the SXWQs. 
Note the very tight relation between $v_{\rm max}$ and $M_V$ for the
SXWQs ($v_{\rm max}\propto L^{0.62\pm 0.08}$), 
as qualitatively expected for some radiation-pressure driven 
outflow scenarios.}
\end{figure}

\clearpage
{\begin{figure}
\plotone{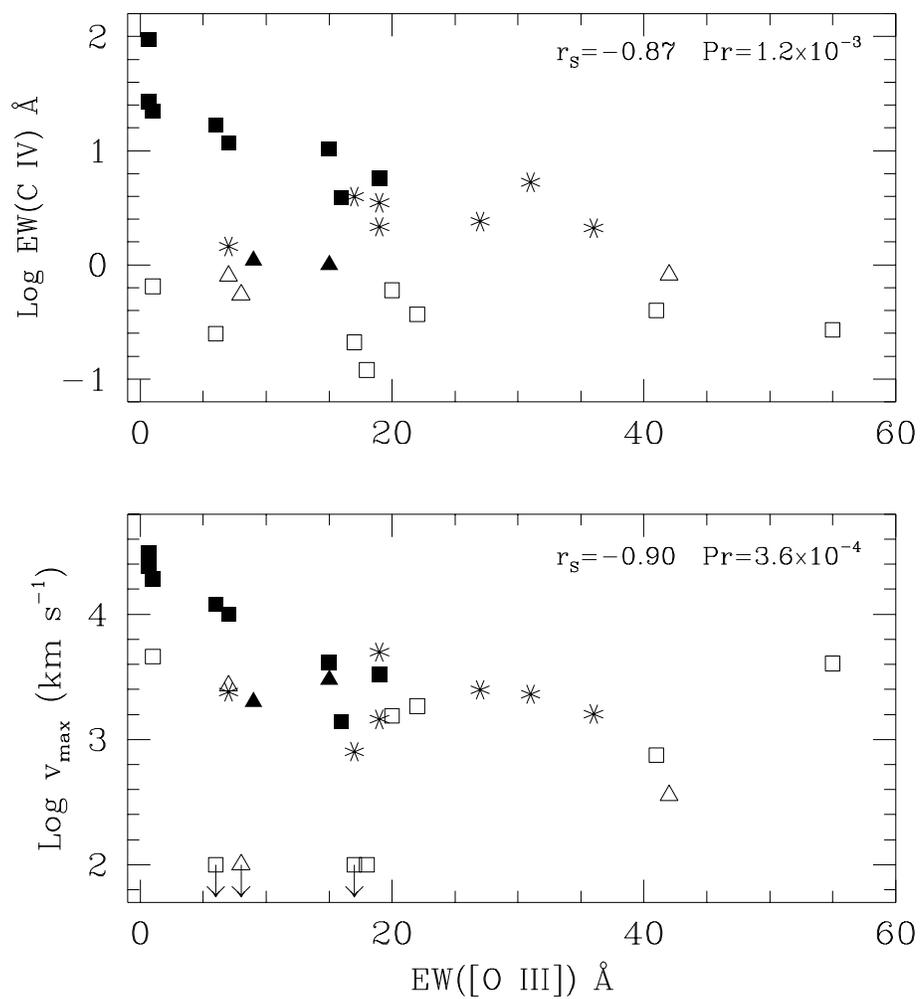}
\caption{The dependence of absorption \CEW\ and $v_{\rm max}$
on \OEW. Both parameters for the SXWQs are significantly 
correlated with \OEW. The values of $r_{\rm S}$ and Pr for
the SXWQs are indicated in each panel.}
\end{figure}

\clearpage
\begin{figure}
\plotone{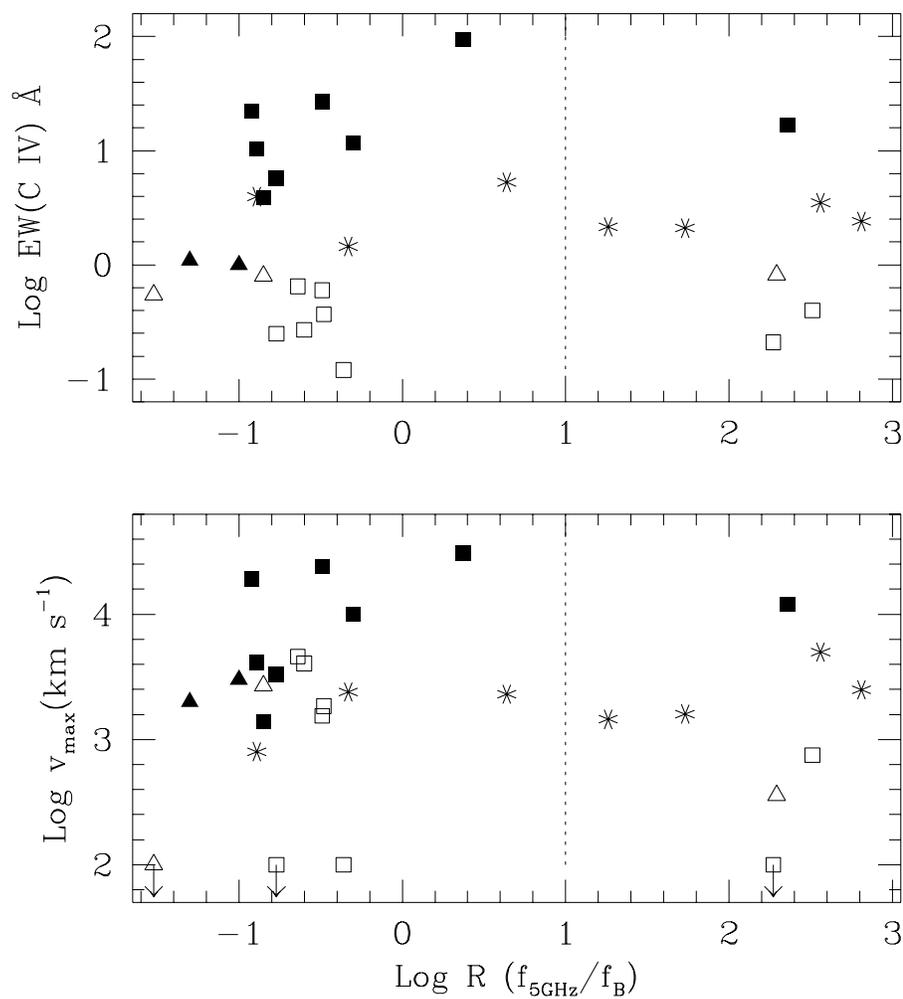}
\caption{The radio-loudness dependence of \CEW\ and of $v_{\rm max}$. 
The vertical dashed line separates radio-quiet and radio-loud AGN.
There may be a hint for somewhat lower \CEW\ and $v_{\rm max}$ in
radio-loud AGN, but the observed differences are not formally
significant (see text).}
\end{figure}

\clearpage
\begin{figure}
\plotone{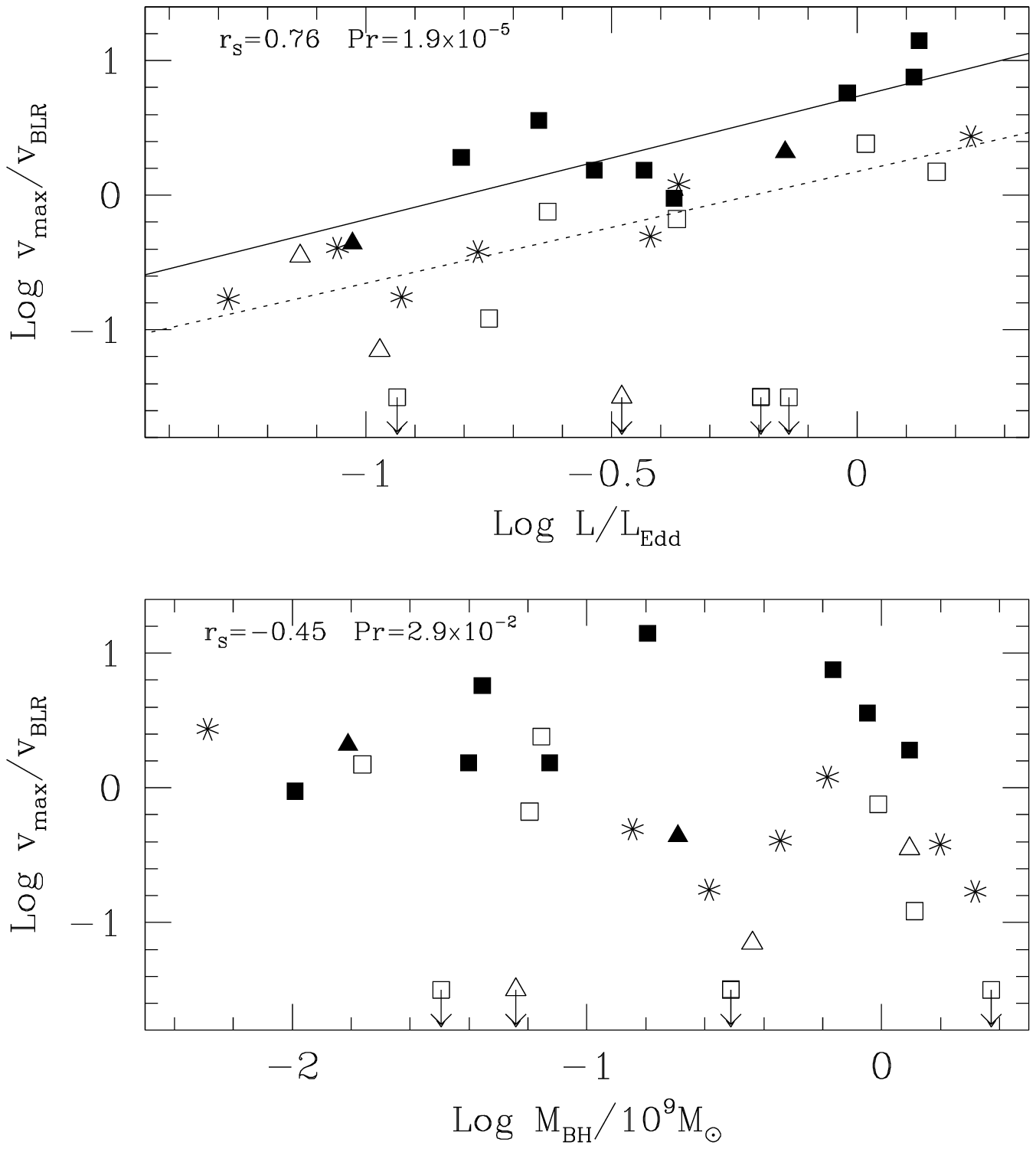}
\caption{The dependence of $v_{\rm max}/v_{\rm BLR}$ on the 
estimated $L/L_{\rm Edd}$ and $M_{\rm BH}$. Note the rather strong correlation 
between $v_{\rm max}/v_{\rm BLR}$ and $L/L_{\rm Edd}$ for all 
absorbed objects, but the only marginal correlation with $M_{\rm BH}$. 
Upper limits represent objects with no significant outflows 
($v_{\rm max}\le 100$~\kms), 
which are excluded from the analysis. The values of $r_{\rm S}$ and
Pr for the 24 AGN with significant outflows 
are indicated in each panel. The solid line is a least-squares fit 
for the 10 SXWQs, which gives 
$v_{\rm max}/v_{\rm BLR}\propto (L/L_{\rm Edd})^{0.91\pm 0.24}$, and the 
dotted line
is a fit for the 14 non-SXWQs, which gives 
$v_{\rm max}/v_{\rm BLR}\propto (L/L_{\rm Edd})^{0.83\pm 0.16}$.}
\end{figure}


\begin{references}
\reference{} 
Arav, N.\ 1997, in ASP Conf.~Ser.~128: Mass Ejection from Active Galactic 
Nuclei, eds. N. Arav, I. Shlosman, and R. J. Weymann, (ASP Press: 
San Francisco), 208
\reference{} 
Arav, N., Li, Z., \& Begelman, M.~C.\ 1994, \apj, 432, 62 
\reference{} 
Arav, N. et al. 2001, ApJ, 561, 118
\reference{} 
Becker, R.~H., White, R.~L., Gregg, M.~D., Brotherton, M.~S., 
Laurent-Muehleisen, S.~A., \& Arav, N.\ 2000, \apj, 538, 72 
\reference{} 
Becker, R.~H.~et al.\ 2001, \apjs, 135, 227 
\reference{} 
Begelman, M., de Kool, M., \& Sikora, M.\ 1991, \apj, 382, 416 
\reference{}
Boller, T., Brandt, W.~N., \& Fink, H.\ 1996, A\&A, 305, 53 
\reference{}
Boroson, T. A. 2002, \apj, in press (astro-ph/0109317)
\reference{}
Boroson, T. A., \& Green, R. F. 1992, \apjs, 80, 109 (BG92)
\reference{}
Boroson, T. A., \& Meyers, K. A. 1992, \apj, 397, 442
\reference{}
Brandt, W. N., \& Boller, Th. 1999, in Structure and Kinematics
of Quasar Broad Line Regions, eds. Gaskell, C. M. Brandt, W. N., 
Dietrich, M., Dultzin-Hacyan, D., Eracleous, M. (ASP Press:
San Francisco), 265
\reference{}
Brandt, W. N., Laor, A., \& Wills, B. J. 2000, \apj, 528, 637 (BLW)
\reference{}
Brotherton, M.~S., Tran, H.~D., Becker, R.~H., Gregg, M.~D., 
Laurent-Muehleisen, S.~A., \& White, R.~L.\ 2001, \apj, 546, 775 
\reference{}
Crenshaw, D.~M., Kraemer, S.~B., Boggess, A., Maran, S.~P., Mushotzky, 
R.~F., \& Wu, C.\ 1999, \apj, 516, 750 
\reference{} 
Ferrarese, L., Pogge, R.~W., Peterson, B.~M., Merritt, D., Wandel, 
A., \& Joseph, C.~L.\ 2001, \apj, 555, L79 
\reference{}
Gallagher, S.~C., Brandt, W.~N., Sambruna, R.~M., Mathur, S., 
\& Yamasaki, N.\ 1999, \apj, 519, 549 
\reference{} 
Gallagher, S. C., Brandt, W. N., Laor, A., Elvis, M., Mathur, S., 
Wills, B. J., \& Iyomoto, N. 2001, ApJ 546, 795
\reference{}
Ganguly, R., Bond, N.~A., Charlton, J.~C., Eracleous, M., Brandt, 
W.~N., \& Churchill, C.~W.\ 2001, \apj, 549, 133
\reference{}
Gebhardt, K.~et al.\ 2000, \apj, 543, L5 
\reference{} 
George, I.~M., Nandra, K., Laor, A., Turner, T.~J., Fiore, F., 
Netzer, H., \& Mushotzky, R.~F.\ 1997, \apj, 491, 508 
\reference{}
George, I.~M., Turner, T.~J., Yaqoob, T., Netzer, H., Laor, A., 
Mushotzky, R.~F., Nandra, K., \& Takahashi, T.\ 2000, \apj, 531, 52 (G00)
\reference{}
Goodrich, R.~W.\ 1997, \apj, 474, 606 
\reference{}
Green, P.~J.~et al.\ 1995, \apj, 450, 51 
\reference{}
Green, P. J. \& Mathur, S.\ 1996, \apj, 462, 637 
\reference{}
Green, P. J., Aldcroft, T. L., Mathur S., Wilkes, B. J., Elvis, M.
2001, ApJ, in press 
\reference{}
Hall, P., 2001, in Mass Outflow in Active Galactic Nuclei: 
New Perspectives, eds. D. M. Crenshaw, S. B. Kraemer, \& I. M.
George, in press (astro-ph/0107182)
\reference{}
Hamann, F. 1998, \apj, 500, 798
\reference{}
Hamann, F., Barlow, T.~A., \& Junkkarinen, V.\ 1997, \apj, 478, 87 
\reference{} 
Jannuzi, B.~T.~et al.\ 1996, \apj, 470, L11
\reference{}
Jannuzi, B.~T.~et al.\ 1998, \apjs, 118, 1 (J98)
\reference{}
Kaspi, S., Smith, P.~S., Netzer, H., Maoz, D., Jannuzi, B.~T.,
\& Giveon, U.\ 2000, \apj, 533, 631
\reference{}
Kaspi, S., Brandt, W. N., Netzer, H., George, I. M., Chartas, G.,
Behar, E., Sambruna, R. M., Garmire, G. P., \& Nousek, J. A.
2001, \apj, 554, 216
\reference{}
Kendall, M., \& Stuart, A. 1977, The Advanced Theory of Statistics, Vol.\ 2
(4th ed.; New York: Macmillian)
\reference{}
Kopko, M., Turnshek, D.~A., \& Espey, B.~R.\ 1994, IAU Symp.~159:
Multi-Wavelength Continuum Emission of AGN, 159, 450 
\reference{}
Korista, K. T. et al. 1992, \apj, 401, 529
\reference{}
Kraemer, S. B., et al. 2001, \apj, 551, 671
\reference{}
Krolik, J.~H.\ 2001, \apj, 551, 72 
\reference{}
Krolik, J.~H.~\& Voit, G.~M.\ 1998, \apj, 497, L5 
\reference{}
Kuraszkiewicz, J., Wilkes, B. J., Brandt, W. N., \& Vestergaard, M. 
2000, \apj, 542, 631
\reference{}
Kwan, J. 1990, ApJ, 353, 123
\reference{}
Laor, A.\ 1998, \apj, 505, L83
\reference{}
Laor, A.\ 2000, NewA Rev, 44, 503
\reference{}
Laor, A.~\& Draine, B.~T.\ 1993, \apj, 402, 441 
\reference{} 
Laor, A., Fiore, F., Elvis, M., Wilkes, B.~J., \& McDowell, J.~C.\ 
1997, \apj, 477, 93 (L97)
\reference{}
Mathur, S.~et al.\ 2000, \apj, 533, L79 
\reference{}
Menou, K., et al. 2001, ApJ, 561, 645
\reference{}
Mineo, T.~et al.\ 2000, A\&A, 359, 471 (M00)
\reference{}
Murray, N.~\& Chiang, J.\ 1995, \apj, 454, L105  
\reference{} 
Murray, N., Chiang, J., Grossman, S.~A., \& Voit, G.~M.\ 1995, \apj, 451, 
498 
\reference{}
Netzer, H., \& Laor, A. 1993, \apj, 404, L51
\reference{}
Netzer, H. 1990 in Active Galactic Nuclei, ed. T. J.L. Courvoisier
\& M. Mayor (Berlin: Springer), 57
\reference{}
Neugebauer, G., et al. 1987, \apjs, 63, 615
\reference{}
Peterson, B.~M.~\& Wandel, A.\ 2000, \apj, 540, L13
\reference{}
Richards, G.~T., York, D.~G., Yanny, B., Kollgaard, R.~I., Laurent-Muehleisen, 
S.~A., \& vanden Berk, D.~E.\ 1999, \apj, 513, 576 
\reference{} 
Risaliti, G., Marconi, A., Maiolino, R., Salvati, M., \& Severgnini, 
P.\ 2001, A\&A, 371, 37 
\reference{} 
Sambruna, R.~M., Eracleous, M., \& Mushotzky, R.~F.\ 1999, \apj, 526, 60 
\reference{} 
Savage, B.~D.~et al.\ 2000, \apjs, 129, 563 
\reference{}
Schmidt, M., \& Green, F. G. 1983, \apj, 269, 352
\reference{}
Scoville, N., \& Norman, C. 1995, \apj, 451, 510
\reference{}
Turnshek, D. A., Monier, E., Sirola, C. J., \& Espey, B. R. 1997,
\apj, 476, 40
\reference{}
Tripp, T.\ M., Lu, L., \& Savage, B.\ D.\ 1998, \apj, 508, 200 
\reference{} 
Wang, T., Brinkmann, W., \& Bergeron, J.\ 1996, A\&A, 309, 81
\reference{} 
Wang, T.~G., Brinkmann, W., Matsuoka, M., Wang, J.~X., \& Yuan, W.\ 
2000, \apj, 533, 113
\reference{} 
Wills, B.~J., Laor, A., Brotherton, M.~S., Wills, D., 
Wilkes, B.~J., Ferland, G.~J., \& Shang, Z.\ 1999, \apj, 515, L53 
\reference{}
Weymann, R.~J., Turnshek, D.~A., \& Christiansen, W.~A.\ 1985, in
Astrophysics of Active Galaxies and Quasi-Stellar Objects, ed. J. S.
Miller (Mill Valley, CA: University Science Books), 333 
\reference{}
Weymann, R.\ J., Morris, S. L., Foltz, C. B., \& Hewett, P. C.
1991, \apj, 373, 23 
\reference{}
Zheng, W., et al. 2001, \apj, 562, 152
\end{references}
\end{document}